\documentclass[12pt,preprint]{aastex}
\newcommand{\om}{\ensuremath{\Omega_m}}

\newcommand{\ola}{\ensuremath{\Omega_{\Lambda}}}
\newcommand{\kmsmpc}{{\ensuremath{{\rm km~s}^{-1}~{\rm Mpc}^{-1}}}}

\newcommand{\cosmol}[3]{\ensuremath{\om = #1,\;\ola = #2,\;H_0 = #3~\kmsmpc}}
\newcommand{\lcdmparm}{\cosmol{0.27}{0.73}{71}}
\newcommand{\etal}{et~al.\/}

\newcommand{\ha}{H\ensuremath{\alpha}}
\newcommand{\hb}{H\ensuremath{\beta}}
\newcommand{\lsun}{\ensuremath{L_\sun}}
\newcommand{\msun}{\ensuremath{M_\sun}}
\newcommand{\msunpyr}{M\ensuremath{_\odot}\,yr\ensuremath{^{-1}}}

\newcommand{\ie}{i.\,e.\,}
\newcommand{\eg}{e.\,g.\,}

\newcommand{\esca}{ergs s\ensuremath{^{-1}} cm\ensuremath{^{-2}} \AA\ensuremath{^{-1}}}

\newcommand{\esc}{ergs s\ensuremath{^{-1}} cm\ensuremath{^{-2}}}
\newcommand{\es}{ergs s\ensuremath{^{-1}}}
\newcommand{\kms}{\mbox{\,km\,s\ensuremath{^{-1}}}}
\newcommand{\cms}{\mbox{\,cm\,s\ensuremath{^{-1}}}}
\newcommand{\HI}{\ion{H}{1}}
\newcommand{\hi}{\ion{H}{1}}
\newcommand{\HII}{\ion{H}{2}}
\newcommand{\iraf}{{\sc iraf}}
\newcommand{\kp}{\ensuremath{K^\prime}}
\newcommand{\ionpar}{\mathcal{U}}
\newcommand{\hrs}{\mbox{$^{\mathrm h}$}}
\newcommand{\mns}{\mbox{$^{\mathrm m}$}}
\newcommand{\scs}{\mbox{$^{\mathrm s}$}}
\newcommand{\dec}[3]{\ensuremath{{#1}\arcdeg{#2}\arcmin{#3}\arcsec}}
\newcommand{\ra}[3]{\ensuremath{{#1}\hrs{#2}\mns{#3}\scs}}
\newcommand{\chisq}{\ensuremath{\chi^2}}
\newcommand{\rchisq}{\ensuremath{\chi^2_{\nu}}}
\newcommand{\mjbks}{mJy\,beam\ensuremath{^{-1}}\,km\,s\ensuremath{^{-1}}}

\shorttitle{Minkowski's Object}
\shortauthors{Croft \etal}

\begin{document}

\title{Minkowski's Object: A Starburst Triggered by a Radio Jet, Revisited}

\author{Steve Croft\altaffilmark{1,2}, Wil van Breugel\altaffilmark{1,2}, Wim de Vries\altaffilmark{1,3}, Mike Dopita\altaffilmark{4}, Chris Martin\altaffilmark{5}, Raffaella Morganti\altaffilmark{6,9}, Susan Neff\altaffilmark{7}, Tom Oosterloo\altaffilmark{6,9}, David Schiminovich\altaffilmark{8}, S.~A.~Stanford\altaffilmark{1,3}, Jacqueline van Gorkom\altaffilmark{8}}\altaffiltext{1}{Institute of Geophysics and Planetary Physics, Lawrence Livermore National Laboratory L-413, 7000 East Avenue, Livermore, CA 94550; scroft@igpp.ucllnl.org}
\altaffiltext{2}{University of California, Merced, P.O. Box 2039, Merced, CA 95344}
\altaffiltext{3}{University of California, Davis, 1 Shields Ave, Davis, CA 95616}
\altaffiltext{4}{Research School of Astronomy and Astrophysics, Australian National University, Cotter Road, Weston Creek, ACT 2611, Australia}
\altaffiltext{5}{California Institute of Technology, Pasadena, CA 91125}
\altaffiltext{6}{Netherlands Foundation for Research in Astronomy, Postbus 2, 7990 AA Dwingeloo, Netherlands}
\altaffiltext{7}{NASA Goddard Space Flight Center, Laboratory for Astronomy and Solar Physics, Code 681, Greenbelt, MD 20771}
\altaffiltext{8}{Department of Astronomy, Columbia University, 538 West 120th Street, New York, NY 10027}
\altaffiltext{9}{Kapteyn Astronomical Institute, University of Groningen, P.O. Box 800, 9700 AV Groningen, The Netherlands}

\begin{abstract}
We present neutral hydrogen, ultraviolet, optical and near-infrared
imaging, and optical spectroscopy, of Minkowski's
Object (MO), a star forming peculiar galaxy near NGC\,541. The observations
strengthen evidence that star formation in MO was triggered by the radio jet
from NGC\,541. Key new results are the discovery of a $4.9 \times 10^8$\,\msun\
double \HI\ cloud straddling the radio jet downstream from MO, where the jet
changes direction and decollimates; strong detections of MO, also showing
double structure, in UV and \ha; and numerous \HII\ regions and associated
clusters in MO. In UV, MO resembles the radio-aligned, rest-frame UV
morphologies in many high redshift radio galaxies (HzRGs), also thought to be
caused by jet-induced star formation. MO's stellar population is
dominated by a 7.5\,Myr-old, $1.9 \times 10^7$\,\msun\ instantaneous burst,
with current star formation rate 0.52\,\msunpyr (concentrated upstream from
where the \HI\ column density is high). This is unlike the jet-induced star
formation in Centaurus~A, where the jet interacts with pre-existing cold gas;
in MO the \HI\ may have cooled out of a warmer, clumpy intergalactic or
interstellar medium as a result of jet interaction, followed by collapse of 
the cooling clouds and subsequent star formation (consistent
with numerical simulations). Since the radio source that
triggered star formation in MO is much less luminous, and
therefore more common, than powerful HzRGs, and
because the environment around MO is not particularly special in terms
of abundant dense, cold gas, jet-induced star formation
in the early universe might be even more prevalent than previously thought.
\end{abstract}


\section{Introduction}

In a previous paper, twenty years ago, \citet{wvb:85} hypothesised
that the starburst seen in the peculiar ``Minkowski's Object''
\citep[MO;][]{mink} in the cluster Abell~194 was triggered by the
radio jet emerging from the nucleus of the nearby active galaxy
NGC\,541. Only one other example of jet-induced star formation was
known at the time, in Centaurus A  
\citep{graham:81,brodie:83}, and the
phenomenon of jet-induced star formation was considered more of a
curiosity than something worthy of general interest. In both cases the
radio galaxies are at low redshift and of relatively low radio
luminosity. Subsequent observations, with more sensitive detectors,
larger telescopes, and better spatial resolution have shown that
jet-induced star formation is more common than initially thought and
examples have now been found in sources at low as well as high
redshift, and with low (FR-I type) or high (FR-II type) radio
luminosity \citep{fr}.

For example, we now know that star formation in low redshift, low
radio luminosity radio galaxies can occur in a variety of environments
such as the central intergalactic medium (IGM) in X-ray clusters like 
Abell~1795 \citep{mcnamara:02,odea:04} and the interstellar
medium (ISM) of their parent galaxies \citep[\eg,][]{wills:04}.  Some
of the most dramatic jet-induced star formation, with star formation
rates (SFR) as high as 1000\,\msunpyr, is associated with very
luminous radio galaxies at redshifts up to $z \sim 4$ \citep{dey:97,bicknell:00}. In all these
cases the radio sources are embedded in relatively dense, and
presumably clumpy, media as evidenced by X-ray halos, cold gas (\HI),
dust, bright line emission or a combination of these. Theoretical
considerations suggest that the star formation may be induced by
moderately slow ($< 1000$\,\kms) radiative shocks as they encounter
IGM or ISM clouds \citep{deyoung:89,rees:89}. These qualitative models
have been developed primarily for the powerful high redshift radio
galaxies, where the star formation is triggered by sideways expanding,
radiative shocks in the relatively dense IGM / ISM of the gaseous
halos in proto-clusters or young parent galaxies
\citep{bicknell:00,klamer:04}.

However, little is known with certainty of the astrophysical
conditions within the radio jets, including the composition of the jet
fluid and its speed, as well as their environment, especially at high
redshift. Better knowledge about this would be of great general
interest, as it would help with modeling outflows from black hole
accretion disks and the conditions of the ISM / IGM in forming
galaxies and their proto-clusters.  Numerical simulations are being
used to explore the physical parameter range that is thought to be
relevant to gain further insight. To help constrain such simulations
MO is of special interest since in this case there is no obvious
indication that the jet is propagating through some especially dense
IGM or ISM environment. The Abell~194 cluster does not have a luminous
``cooling'' X-ray halo like Abell~1795, and there is no evidence for
much cold gas outside the jet like in Centaurus~A.  Thus a better
understanding of how the star formation in MO was triggered by the jet
would also help in determining how much of a general phenomenon this
might be.

So why did the NGC\,541 radio jet trigger star formation? Was it just
a chance encounter with a gas rich galaxy that wandered into path of
the radio jet, which was then ``rejuvenated''? This would have been a
lucky shot, given that MO is the only peculiar system within the
central $\sim 1$,Mpc of Abell~194 and that its total (projected) diameter
is the same as that of the radio jet. However, if there were an old
stellar population underlying the starburst in MO, then this could not
be ruled out. On the other hand, as the observations of Abell~1795
have shown, the interaction of jets with warm, clumpy gas in cooling
flow clusters may well trigger the collapse of IGM clouds and
subsequent star formation.

The development of the new massively-parallel {\sc cosmos} numerical
simulations package \citep{cosmos} at Livermore allowed \citet{mo_sim}
to simulate the collision of an FR-I type jet with a gas cloud in an
effort to see whether star formation could be induced by the jet in
the case of MO. They assumed that NGC\,541 is surrounded by a
multi-phase medium resembling galaxy cluster atmospheres
\citep[\eg,][]{ferland:02}; the simulations began with a hot
``mother'' cloud with a semimajor axis of 10\,kpc, a semiminor axis of
5\,kpc, density $0.1$\,cm$^{-3}$, and temperature $10^6$\,K. Within
the mother cloud, denser, warm clouds were assumed to be embedded with
typical sizes of 100\,pc, densities of $10$\,cm$^{-3}$, and
temperatures of $10^4$\,K. Studies of the proto-typical FR-I radio
galaxy 3C\,31 \citep{laing:02} were used to estimate plausible
parameters for the jet at the location of MO, 15\,kpc from NGC\,541: a
velocity of $9 \times 10^4$\,\kms, a density of $10^{-4}$\,cm$^{-3}$,
and a temperature of $10^9$\,K. \citet{mo_sim} concluded that it was
indeed plausible that such star formation could be triggered by the
driving of a radiative shock into such a cloud, causing the cloudlets
to compress, radiatively cool, and break up into numerous dense, cold
fragments. These fragments survive for many dynamical timescales and
are presumably precursors to star formation.

Motivated by these results from numerical simulations, and by the
availability of new observing facilities since the previous study, we
find this an opportune time to revisit MO in an attempt to better
understand the nature of jet / cloud interactions.  We will first
discuss the various new observations obtained, supplemented by
archival observations. Then we proceed with an analysis of the
spectroscopic and spectral energy distribution information, a
discussion of the kinematics, morphology and time scales, and finally
our conclusions and ideas for future work. We assume an \lcdmparm\
cosmology \citep{wmap}. At $z = 0.019$, MO and the Abell~194 cluster are
both at a comoving radial distance of 78.7\,Mpc, and have a
linear scale size of 0.374\,kpc~arcsec$^{-1}$.

\section{Observations}

\subsection{VLA radio continuum data}

\citet{wvb:85} published 1.4\,GHz Very Large Array (VLA) maps (taken in A-, B-,
and C-array) of MO. These original radio data were re-reduced for the
present investigation using standard routines in the AIPS software
package. This was done in order to benefit from the cumulative
improvements to the software over the years since 1985. This new radio
continuum image is included in Figs.~\ref{fig:hstrad} and
\ref{fig:pfcam}, and has a resolution of 3\arcsec. We refer to the original paper for more details on
the radio data.

\subsection{HST archival $V$ and $I$-band WFPC2 images}

Although not the main target, MO was within the field of two short
(460\,s each) Hubble Space Telescope (HST) WFPC2/F555W and WFPC2/F814W
observations of NGC\,541 \citep{mo:hst}, which are available in the
archives. These are the highest spatial resolution images of MO, with
a measured FWHM of 0.17\arcsec\ at F555W and 0.20\arcsec\ at F814W.
\clearpage
\begin{figure}
\centering
\includegraphics[width=\hsize]{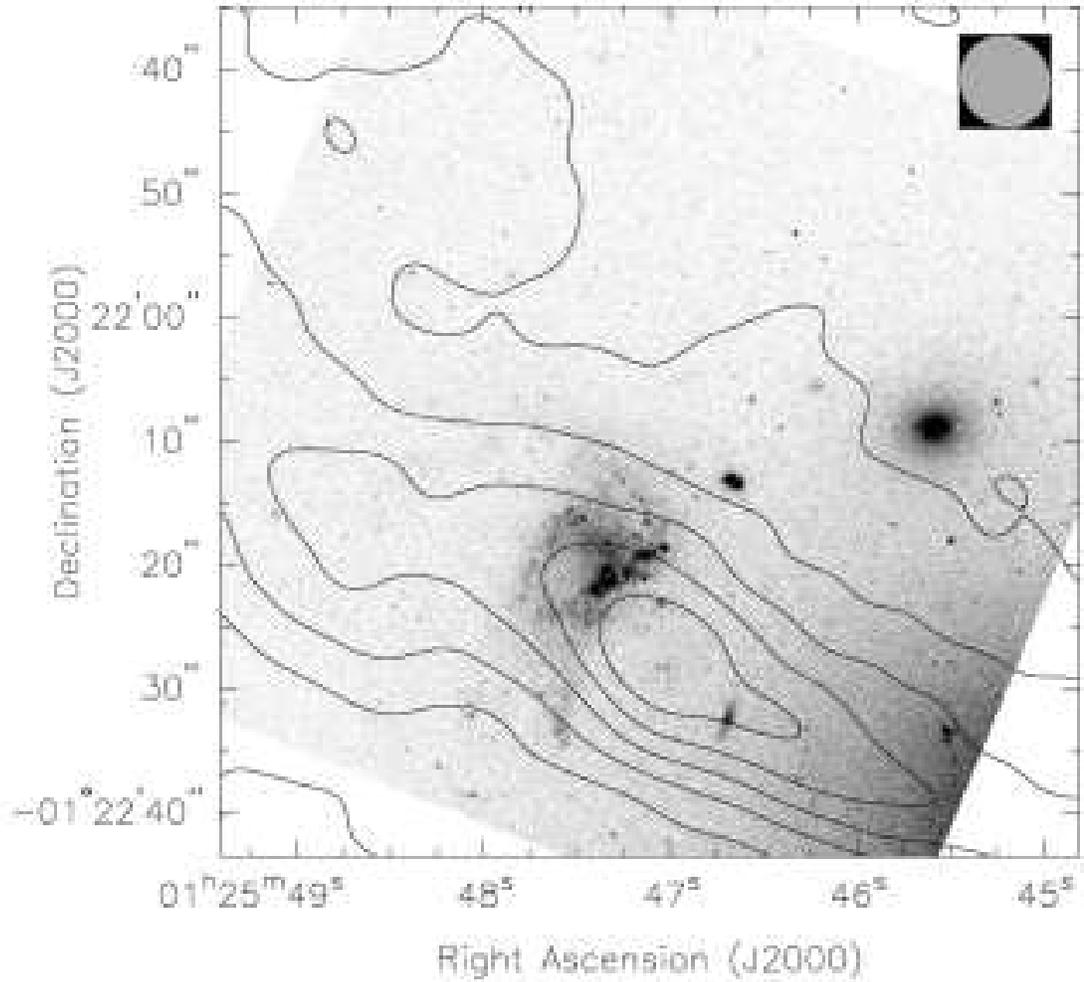}
\caption{\label{fig:hstrad}Coadded HST F555W / F814W image of MO, overlaid with radio continuum contours from the VLA. Contours are from 0 to 10\,mJy\,beam$^{-1}$ at intervals of 2\,mJy\,beam$^{-1}$. The grey ellipse represents the $7{\farcs}8 \times 7{\farcs}3$ {\sc clean} beam (PA $= 36$\degr) of the radio observations.
% CLEAN BMAJ=  2.1579E-03 BMIN=  2.0222E-03 BPA=  36.17
}
\end{figure}
\clearpage
MO shows considerable structure with a total projected size which is
approximately the same as the jet diameter
(Fig.~\ref{fig:hstrad}). The largest and brightest component
(hereinafter referred to MO-North) is found along the northern edge of
the jet. Its brightest portion, a 3\,kpc-long bar-like feature lies on
the side closest to NGC 541 and extends at almost a right angle to the
jet. A much fainter, linear feature (hereinafter referred to as
MO-South) extends to the south across the jet. The central, high
surface brightness portion of the radio jet bisects these two
components just south of the position of the peak of the star
formation as seen in the optical images (Fig.~\ref{fig:hstrad}). Apart
from the bar, MO-North contains several filamentary structures
downstream in the jet, and is dotted with compact (mostly unresolved
at the $\sim 75$\,pc resolution of HST) young star clusters and \HII\
regions.

\subsection{Lick wide field $R$-band imaging}

\begin{figure*}
\centering
\includegraphics[width=\hsize]{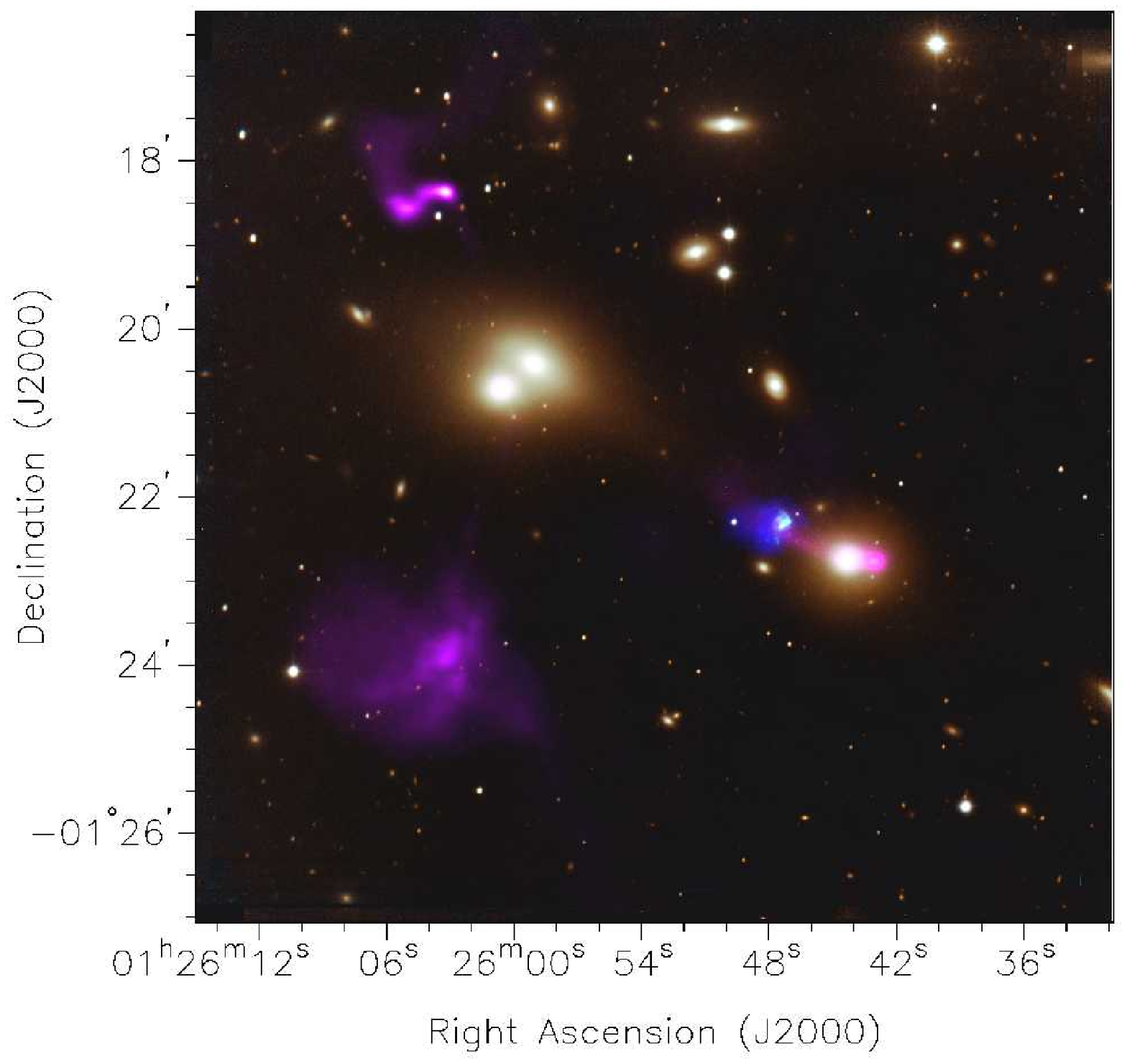}
\caption{\label{fig:pfcam} Coadded PFCam ($BR$) image of the field of
Abell 194 (color), overlaid with radio continuum (purple), \HI\ data
(dark blue), and \ha\ data (light blue). Minkowski's Object is clearly
seen as the peculiar object at \ra{1}{25}{47}, \dec{-1}{22}{20}, forming from the cloud of \HI\ in the path
of the jet from NGC\,541 (\ra{1}{25}{44}, \dec{-1}{22}{46}). The radio galaxy to the left of the image is
3C\,40 (\ra{1}{26}{00}, \dec{-01}{20}{43}), associated with NGC\,547. The galaxy at \ra{1}{25}{59}, \dec{-01}{20}{25}, interacting with NGC\,547, is NGC\,545.  }
\end{figure*}

\begin{figure*}
\centering
\includegraphics[width=\hsize]{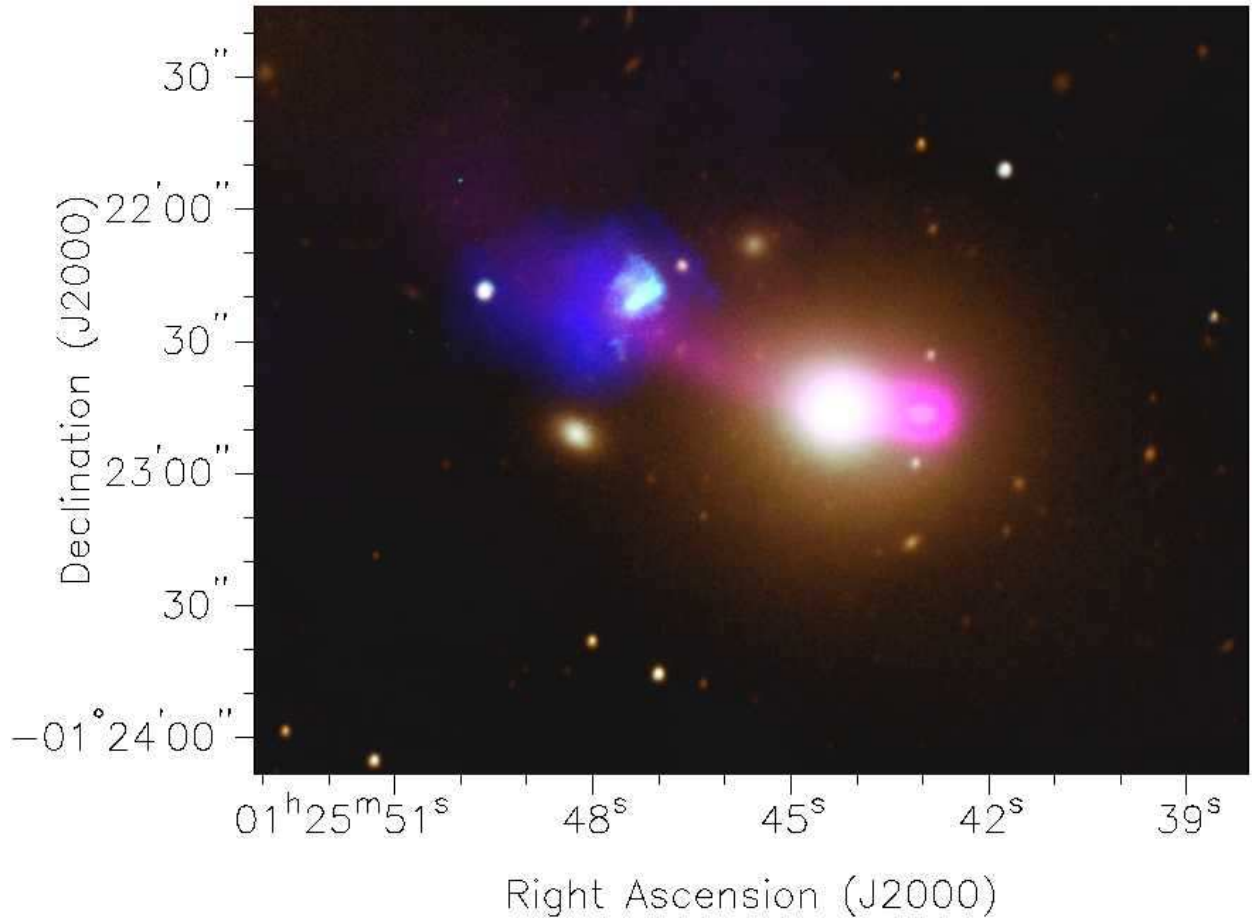}
\caption{\label{fig:pfcamzoom} Zoom of figure~\ref{fig:pfcam} to show in more detail the region surrounding MO. }
\end{figure*}

To investigate the local environment of MO, the central portion of the
Abell~194 cluster was imaged by Michael Gregg on 2002 October 6, using
the Prime Focus Camera (PFCam) on the Lick Observatory Shane 3-m
telescope. Eight 600\,s exposures were made using the Spinrad $R$\
filter, and five 600\,s exposures were made in $B$-band. The frames
were reduced in the standard manner in \iraf, and coadded using
offsets determined from individual astrometric solutions for each
frame. The measured FWHM in the final image is 1.5\arcsec\ and the
total field-of-view is 11\arcmin~$\times$~11\arcmin\
(Figs.~\ref{fig:pfcam} and~\ref{fig:pfcamzoom}).
\clearpage
\clearpage
\clearpage

The image shows a large ($\sim 20 \times 45$\,kpc) stellar bridge
connecting the ellipticals NGC\,541 and the interacting galaxies
NGC\,545 / 547, which are associated with the large FR-I radio source
3C~40 \citep{simkin:76}. MO is located in this bridge, at least as
seen in projection (Fig.~\ref{fig:pfcamr}), opening the possibility
that the gas associated with MO has had its origin in some previous
tidal interaction between these galaxies. The redshifts of these
various systems are consistent: $z = 0.0181$
\citep[NGC\,541;][]{a194z}, $z = 0.0189$ (MO; this paper), $z =
0.0178$ \citep[NGC\,545;][]{a194z}, and $z = 0.0182$
\citep[NGC\,547;][]{n547z}.

\subsection{Narrow-band \ha\ imaging}

To determine the total extent of the star formation in MO and search
for other possible emission-line features we obtained a narrow-band,
dithered \ha\ image on 2004 December 17 using ``Imager'' on the 2.3-m
Advanced Technology Telescope (ATT) at Siding Spring
Observatory. Three 200\,s exposures of MO were made in the $R$-band
filter, and three 600\,s exposures in a narrow-band filter centered at
668\,nm (to match redshifted \ha\ in MO). Three 30\,s exposures in
$R$-band were made of the emission-line flux standard 205-26.7
\citep{dopita:97}, and three 100\,s exposures in a narrow-band filter
centered at 658\,nm to provide the \ha\ flux calibration. The images in
each filter were reduced in the standard manner in \iraf, using sky
flatfields appropriate to each filter. Once the individual images had
been registered and co-added, an astrometric fit was performed for
each coadd, and the broad-band images were registered and
geometrically scaled so that the pixel positions of objects in the
broad-band and narrow-band images matched to sub-arcsecond accuracy.

Continuum subtraction was performed using the technique of
\citet{boker:99}. Briefly, the reduced, registered broad-band and
narrow-band images were rebinned to a size of $135 \times 135$ pixels
(each pixel representing 256 pixels of the original image). The
intensity (in DN per second) of each pixel in the rebinned $R$-band
image was then plotted against that of the same pixel in the rebinned
narrow-band image, excluding the brightest 500 pixels in the
narrow-band image. A linear least-squares fit was performed, the
slope of which represents the scaling factor between the two
filters. The unbinned $R$-band image was then convolved with a
Gaussian filter to match the PSF of the narrow-band image. It was
then scaled by the slope of the fit obtained above, and subtracted
from the unbinned narrow-band image.  

\begin{figure*}
\centering
\includegraphics[width=\hsize]{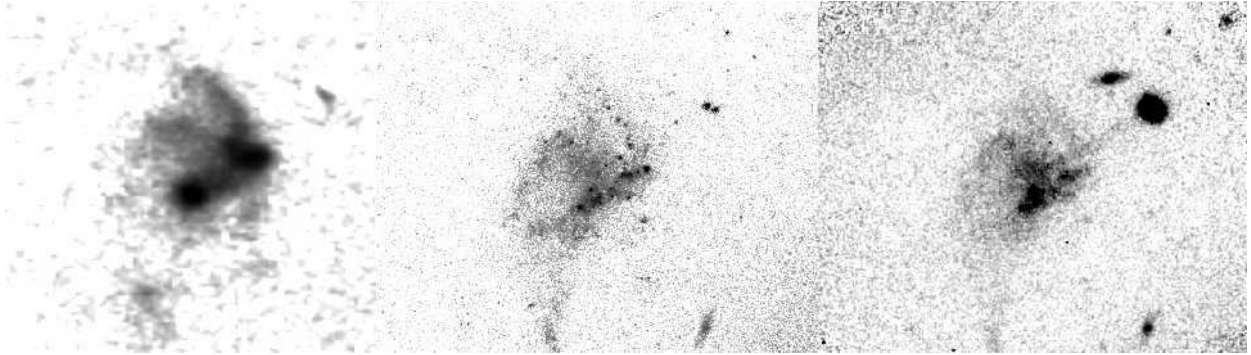}
\caption{\label{fig:haim}{\em Left:} Continuum-subtracted ATT \ha\
emission image of MO. {\em Center:}\label{fig:f555} HST F555W (stellar
continuum) image. {\em Right:}\label{fig:nirc} Keck NIRC \kp\
image. All three images are aligned and show the same 1.0~arcmin$^2$
region of sky.  }
\end{figure*}

The scaling obtained for the standard star observations was 0.0438,
while that for the MO observations was 0.0554. This is presumably due
to differences in the narrow-band filter transmission curves (since
the same $R$-band filter was used for both fields), and the fact that
the 668 filter samples a different part of the SEDs of the objects in
the field to the 658 filter. The latter should be relatively
unimportant due to the relatively small shift in wavelength between
the two filters, except in the case of strong emission lines which may
fall in one filter but not the other (\eg, \ha\ in
particular). Although this is the case for the main targets in each
field, the good fit obtained when all binned pixels are included
suggests that this effect is negligible, and that the main difference
is due to the relative transmission of the filters. Therefore, we
scale the \ha\ fluxes measured from MO by the ratio of the slopes
(\ie, the MO fluxes are multiplied by 0.791) to account for this.
The resulting continuum-free \ha\ image is shown in
Fig.~\ref{fig:haim} and has a resolution of 1.3\arcsec\ FWHM.

\clearpage
\clearpage

The total flux integrated within an equivalent 16\arcsec\ diameter
aperture (after correction for Galactic reddening - see
section~\ref{sec:red}) is determined to be $8.6 \times
10^{-14}$\,\esc, yielding an \ha\ luminosity of $6.6 \times
10^{40}$\,\es. This is approximately twice the value obtained by
\citet{wvb:85} -- some 14\% of this increase is attributable to the
different cosmology used in their paper, and the remainder is
presumably due to improvements in both the sensitivity and calibration
of the current \ha\ observations.

The \ha\ image is similar in its morphological features to the optical
continuum images, with a strong concentration of \HII\ regions and
their associated clusters in the linear bar-like feature facing the
incoming jet, and more diffuse filaments of emission on the downstream
side. These are more conspicuous in \ha\ than in the continuum,
suggesting that the clusters in this region are somewhat younger than
in the bar. MO-South is more clearly defined in \ha. The overall
morphology is highly suggestive of a jet-cloud interaction, with
subsequent star formation triggered by the jet.

\subsection{Keck ESI $R$-band and NIRC $J$ and \kp-band imaging}

To complement the HST $V$ and $I$-band observations, and to provide
spectral energy distribution data over the widest possible range, we
obtained Keck $R$, $J$ and \kp-band images.

\begin{figure}
\centering
\includegraphics[width=\hsize]{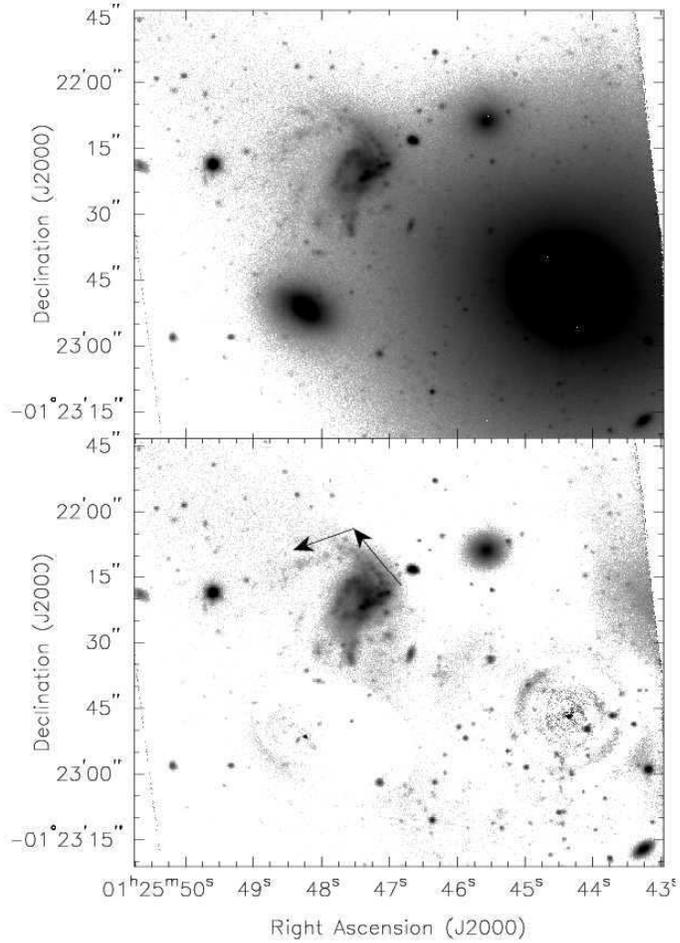}
\caption{\label{fig:esipp} Keck ESI $R$-band image of MO (top) with
elliptical galaxy models subtracted (bottom) as described in the text. Both images are displayed using the same logarithmic scaling. The arrows in the bottom panel illustrate the filamentary feature downstream of the central bar, discussed in section~\ref{sec:filament}. They are not intended to suggest that this feature necessarily formed due to flow of material from the central regions of MO.
}
\end{figure}

In the $R$-band, a 75\,s image centered on MO was obtained with the
Echellette Spectrograph and Imager \citep[ESI;][]{esi} on Keck II on
2002 September 7, and calibrated and reduced in \iraf\ in the standard
manner. The FWHM of the seeing was 0.7\arcsec\ and the image is shown
in Fig.~\ref{fig:esipp}. This image clearly shows the existence of a
fainter structure \label{sec:filament} extending about 30\arcsec\ downstream of the bright bar of MO-North, which was also faintly detected in the \ha\ image.
\clearpage
\clearpage
600\,s exposures in $J$ and \kp\ were obtained with the Near Infrared
Camera (NIRC) on Keck I on 2004 December 17. A 5-position dither, with
dither spacing 5\arcsec\ and 60\,s per exposure ($3 \times 20$\,s in
$J$, $4 \times 15$\,s in \kp), was used. Between each on-source dither
position, a 40\arcsec\ nod to a nearby blank area of sky was made, and
an additional 60\,s exposure off-source was made (yielding the same
dither pattern as for the on-source exposures). The on-source images
were registered and combined. The off-source images were combined with
object masking and without registration to form a sky image, which was
scaled and subtracted from the on-source image. The final reduced
images cover a field of 100\arcsec\ $\times$ 100\arcsec\ with a FWHM
of 0.5\arcsec.

The \kp-band image is shown as the third panel in
Fig.~\ref{fig:nirc}. The \kp-band morphology very closely resembles
the optical F555W image taken with the HST. This shows that there
cannot be a large old stellar population present in MO, and that the
\kp-band light is probably dominated by young supergiants.

\subsection{GALEX imaging}
\label{sec:galex}

The field containing the NGC\,541 / 545 / 547 galaxies in Abell~194 was observed as part of the GALEX  \citep{galex}  Nearby Galaxy Survey \citep{galexngs} in the near-UV (NUV, 1750 -- 2800\,\AA) and far-UV (FUV, 1350 -- 1750\,\AA) passbands on 2003 October 23. Exposure times were 1700\,s in each band. The data were reduced and photometrically calibrated using the GALEX pipeline \citep{galexpipe}. The 5$\sigma$ AB limiting magnitude in both bands is $m \sim 23$ corresponding to a surface brightness sensitivity of $\sim$26.5~mag\,arcsec$^{-2}$ averaged over the PSF. The FWHM in the NUV and FUV images is 5.6\arcsec\ and 4.0\arcsec\ respectively.

MO was prominently detected in both bands and shows similar overall
structure to the optical, \ha\ and near-IR images, even given the
relatively low resolution of GALEX.  The GALEX field of view is large
and covers the central 1.2\degr\ of the Abell~194 cluster, yet MO is
among the brightest resolved UV structures in this entire field.

\begin{figure*}
\centering
\includegraphics[width=0.45\hsize]{f6a.eps}\hspace{0.05\hsize}%
\includegraphics[width=0.45\hsize]{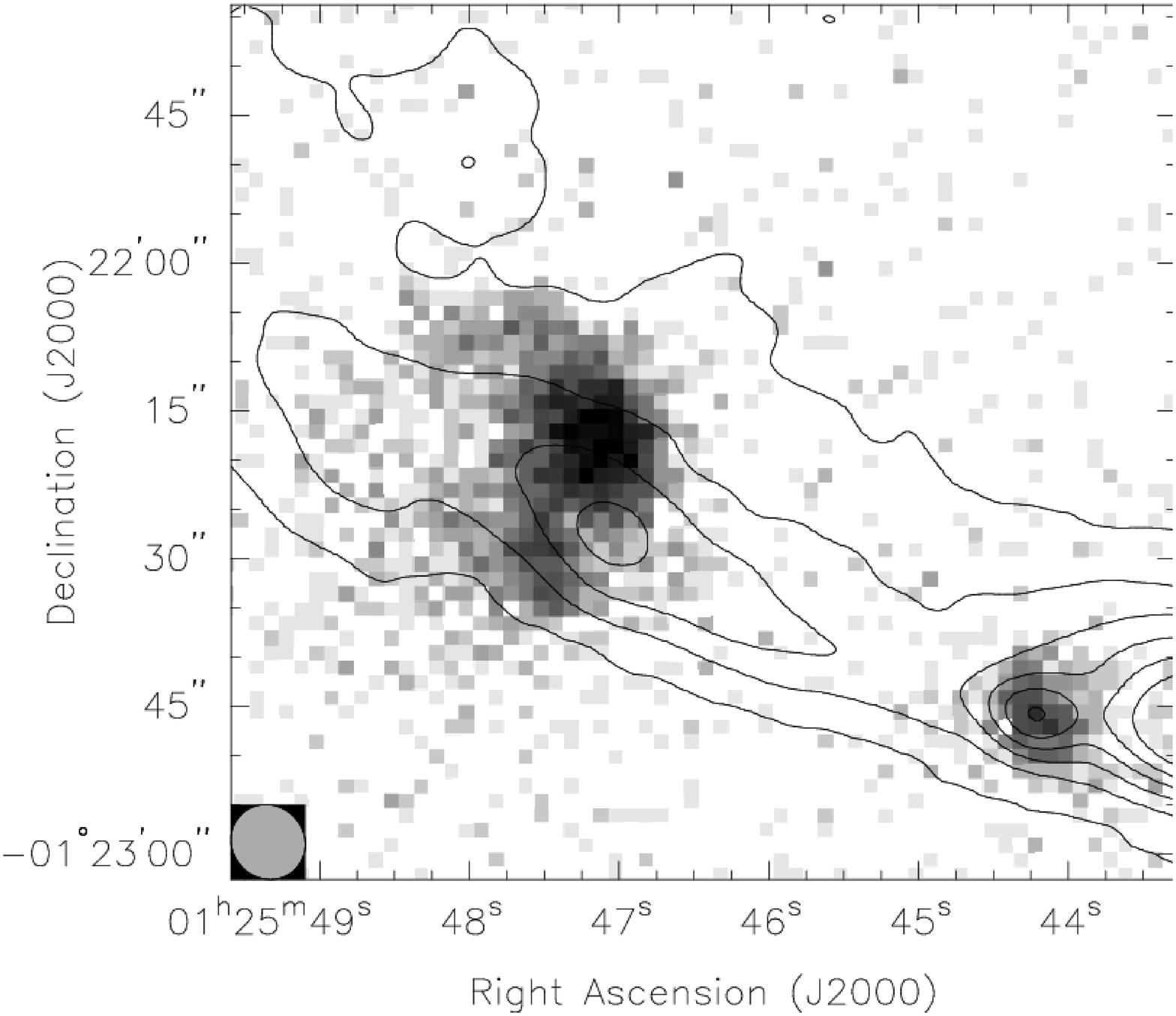}
\caption{\label{fig:fuvcont} {\em Left:} GALEX NUV image, overlaid with \HI\
contours from the VLA. The NUV image is shown with a logarithmic scaling to emphasise faint structures; note that the blob at \ra{01}{25}{50} is a foreground star. \HI\ contours are from 0.5 -- 6.5\,\mjbks, at intervals of 1.0\,\mjbks. The grey ellipse represents the $17{\farcs}9 \times 15{\farcs}4$ {\sc clean} beam (PA $= 30$\degr) of the \HI\ observations. {\em Right:} GALEX FUV image (shown with a logarithmic scaling), overlaid with radio continuum contours from the VLA. Radio contours are from 2.0 -- 17.0\,mJy\,beam$^{-1}$ at intervals of 3.0\,mJy\,beam$^{-1}$. The grey ellipse represents the $7{\farcs}8 \times 7{\farcs}3$ {\sc clean} beam (PA $= 36$\degr) of the radio observations. Note that the two panels are not shown to the same scale; the relationship between the radio jet and the \HI\ is better seen in Fig.~\ref{fig:radhi}.
}
\end{figure*}

\begin{figure}
\centering
\includegraphics[width=\hsize]{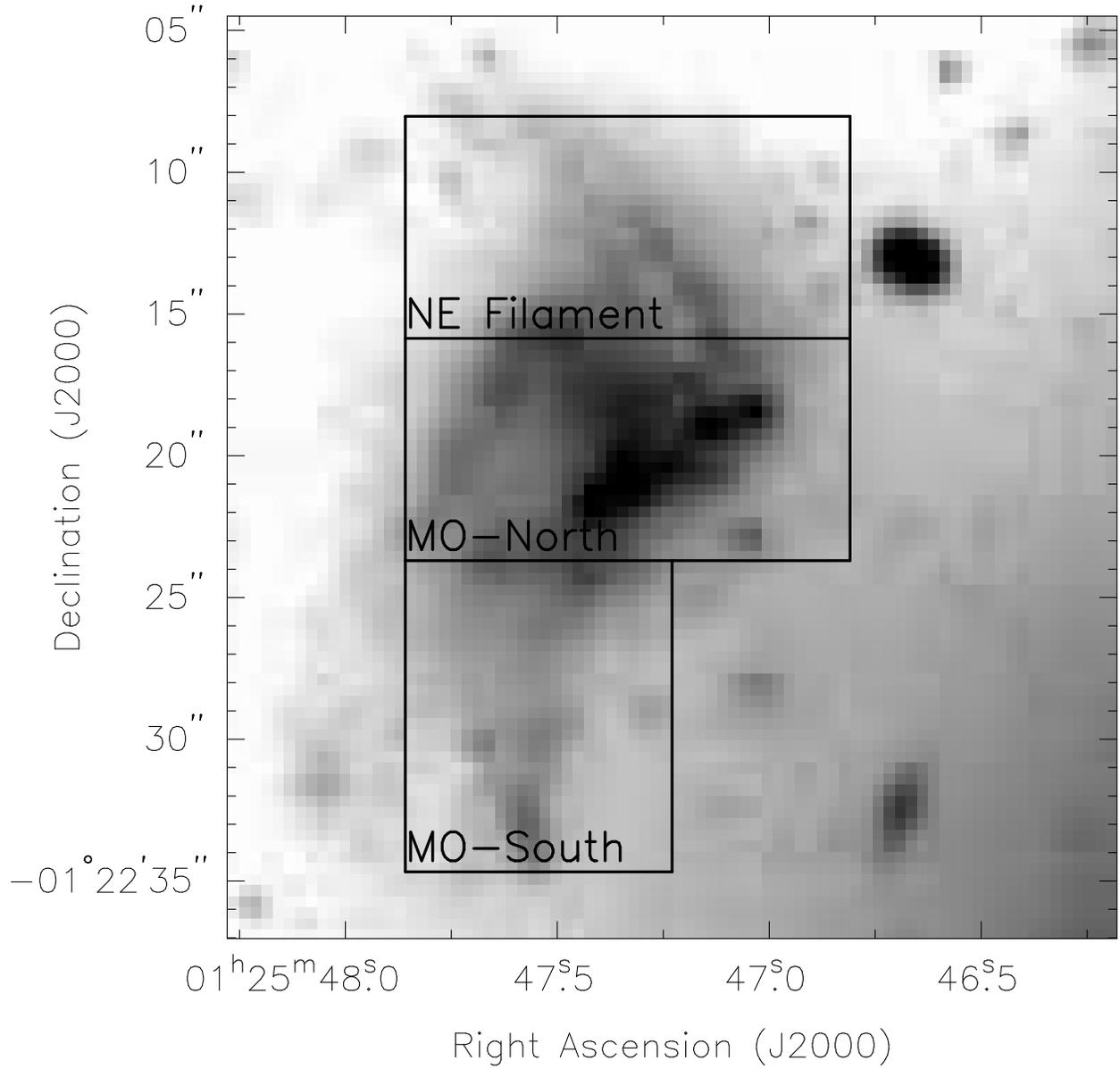}
\caption{\label{fig:threephot} Binned ESI-R image of MO, overlaid with
boxes showing the regions used for the three-aperture photometry (see
text).  }
\end{figure}

The images were overlaid with the VLA radio continuum and \HI\
observations (Section~\ref{sec:hi}), as shown in Fig.~\ref{fig:fuvcont}. In both UV
bands MO appears as a double source, with a clear separation of
MO-North and MO-South straddling the central axis of the radio jet,
with MO-North being the brightest component. Both components are
clearly extended downstream, paralleling the radio jet, as in the
optical and near-IR images.
\clearpage
% A: levels 0.0005 0.0015 0.0025 0.0035 0.0045 0.0055 0.0065
% B: levels 0.002 0.005 0.008 0.011 0.014 0.017
\clearpage
The total observed near-UV flux from MO was determined to be $1.78
\times 10^{-15}$\,\esca, and the far-UV flux was $3.63 \times 10^{-15}$
\esca, yielding values corrected for Galactic extinction of $2.57
\times 10^{-15}$\,\esca\ and $5.11 \times 10^{-15}$ \esca\
respectively. This is in good agreement with a flux of $2.96 \times
10^{-15}$\,\esca\ at 1350\,\AA\ from an archival International Ultraviolet
Explorer spectrum (K.~Chambers, priv.\ comm.).  We also extracted
photometry for three smaller, rectangular apertures, centered on
MO-North, the NE filament downstream from MO-North, and MO-South
(Fig.~\ref{fig:threephot}). These GALEX photometry data allow us to
put strict limits on the stellar age distribution of MO, as we shall
discuss in Section~\ref{sec:timing}.

\clearpage
\clearpage
\subsection{The Radio \HI\ VLA observations}
\label{sec:hi}

To search for cold gas associated with MO and its surroundings, we
performed \HI\ observations with the VLA in its C-array configuration
on 2002 April 2. The total integration time on source was 4.3\,h. A
3\,MHz bandwidth was used, using 128 spectral channels, covering the
heliocentric velocity range 5305\,\kms\ to 5815\,\kms (the full observing band is 600\,\kms, but on either side of this band are a few channels that are not usable). The
data were calibrated following standard procedures using the {\sc miriad}
software package. A spectral-line data cube was made using robust
weighting (robustness = 0.5) giving a spatial resolution of
18\arcsec\ $\times$ 15\arcsec\ (PA = 29\degr).  Hanning
smoothing was applied to the datacube, resulting in a velocity
resolution of 10.3\,\kms. The noise in the final datacube is
0.54\,mJy\,beam$^{-1}$.

\begin{figure}
\centering
\includegraphics[width=\hsize]{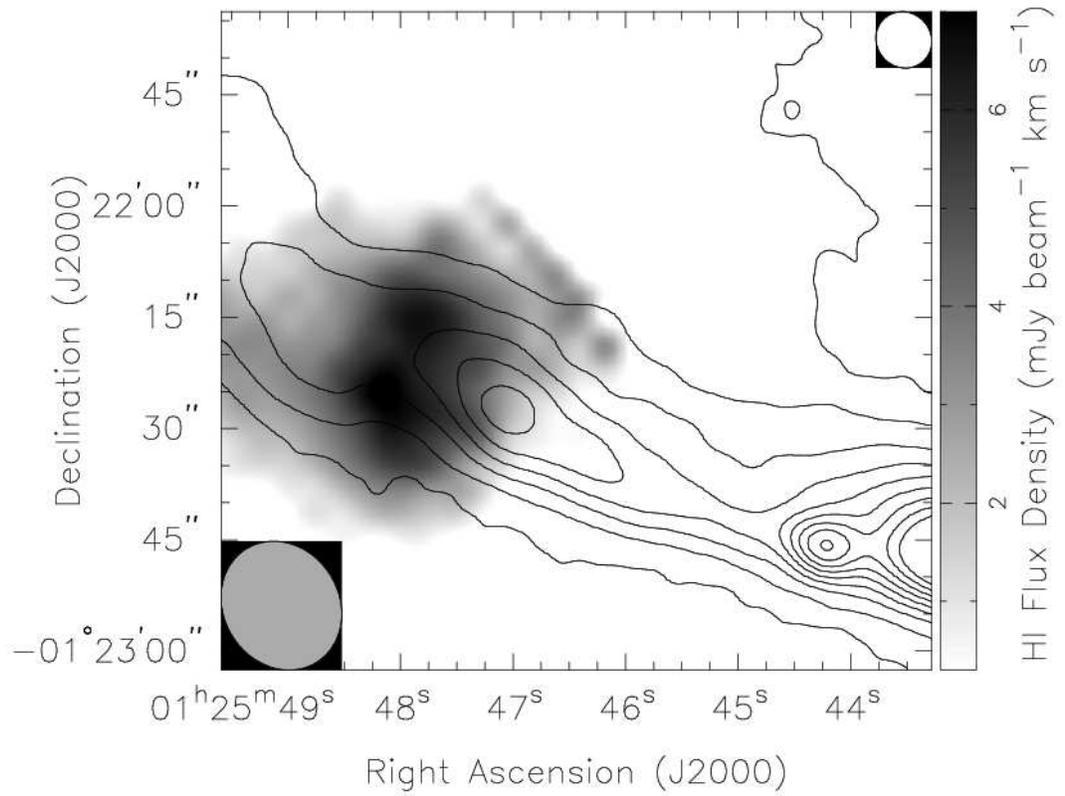}
\caption{\label{fig:radhi} Map of the \HI\ cloud adjoining MO, with
radio continuum contours from the VLA overlaid. The \HI\ cloud is shown with a linear scaling from 0.3 -- 7\,\mjbks, corresponding to the greyscale wedge at right. Continuum contours are from 1 -- 17\,mJy\,beam$^{-1}$ at 2\,mJy\,beam$^{-1}$ intervals. The \HI\ beam is shown as a grey ellipse (lower left), and the continuum beam is shown as a white ellipse (upper right).}
\end{figure}

Within the entire primary beam (0.5\degr) the only \HI\ that we detect
is from MO (Fig.~\ref{fig:radhi}), but this may be due to our
relatively narrow band covering only $\sim 500$\,\kms\ (the velocity range of
the cluster is surely larger than that).  Indeed, a short test
observation with the WSRT in August 2004 of MO with a wider band of 10\,MHz covering 2000\,\kms (using 1024 channels) showed at least 3 other
\HI\ objects, apart from MO, at velocities just outside the band used
with the VLA for the MO observations.  A more detailed observation of
Abell~194 with the WSRT would likely reveal more objects.

\clearpage
\clearpage
The \HI\ emission is dominated by two ``blobs'' sitting on either side
of the jet and is offset from MO, with the brightest \HI\ $\sim
10$\arcsec\ (3.6\,kpc) downstream from the peak of the \ha\ emission
(the main star formation region). The position of these maxima is
coincident with the downstream extensions of MO-North and MO-South
noted in the GALEX image. In addition, two faint \HI\ filaments are
found extended along the edges of the jet. One of these lies to the
north of MO-North -- upstream in the sense of the jet
motion. The other forms a continuation of MO-South downstream. The jet
itself changes direction and decollimates downstream from the \HI\
clouds, which suggests that it is the jet--gas interaction which has
disrupted the jet.

\begin{figure}
\centering
\includegraphics[width=\hsize]{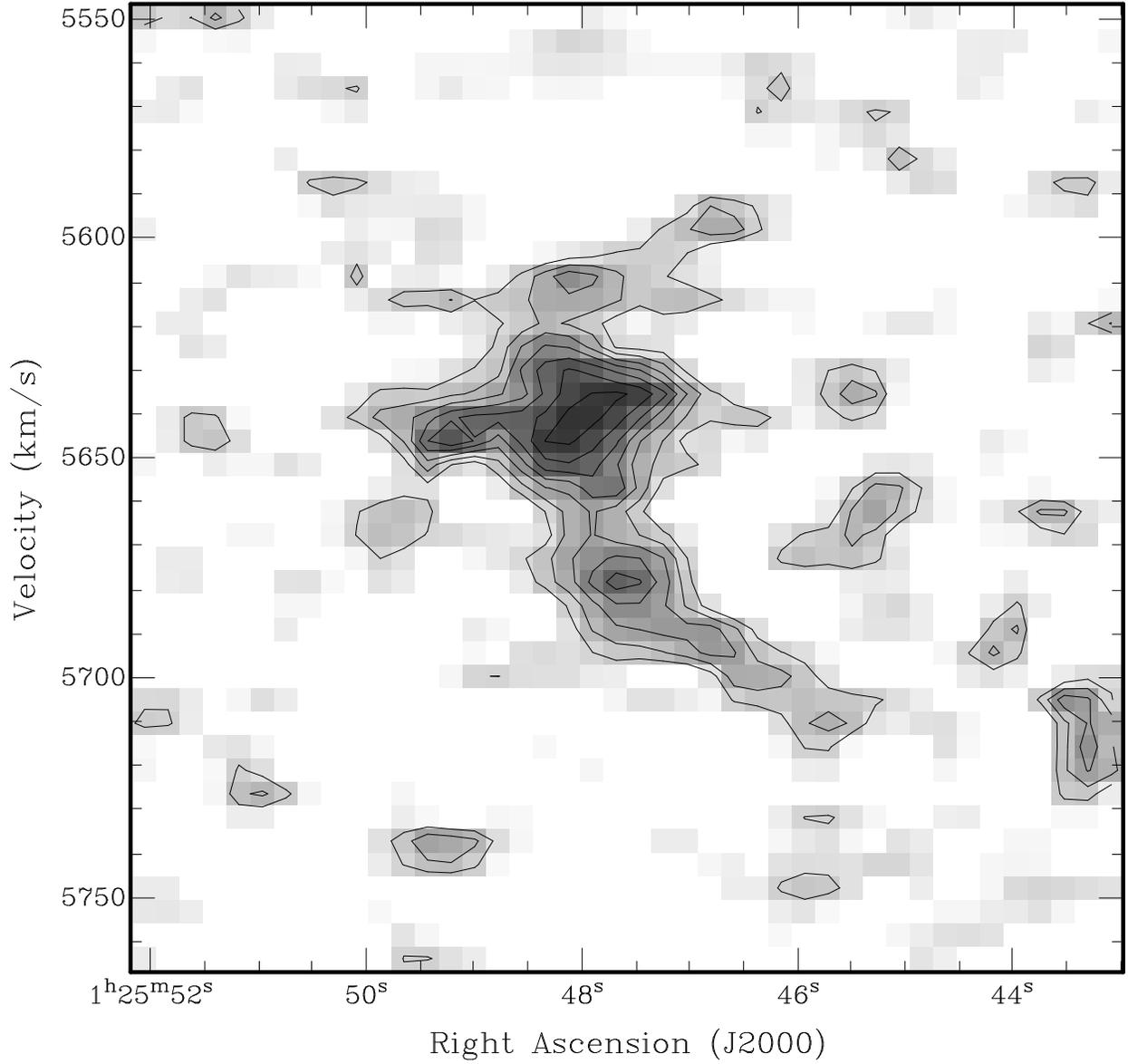}
\caption{\label{fig:hikin} Velocity slice through the MO \HI\ map,
showing velocity shear near the outer boundaries of the \HI\ cloud and
jet. Contour levels are 0.5, 1.0, 1.5, 2.0, 2.5, 3.0 mJy\,beam$^{-1}$, and encode the same information as the greyscale.}
\end{figure}

The \HI\ kinematics, along an east-west slice (see
Fig.~\ref{fig:hikin}) show that most of the \HI\ gas is
relatively quiescent with a average velocity $v_{HI} \sim$
5640\,\kms, or $z = 0.0188$. This is in excellent agreement with the
redshift $z = 0.0189$ of the star-forming region obtained from the
LRIS spectroscopy. To make this plot, the data were integrated in
declination over about 35\arcsec. This was done to show the faint low-
and high-velocity wings. There is a smooth north--south velocity gradient and
the North and South \HI\ blobs have a velocity difference of about $\sim
50$\,\kms.
\clearpage
\clearpage
We derive a total \HI\ mass for MO of $4.9 \times 10^8$\,\msun.
This was done assuming the \HI\ is optically thin.  Then the mass follows from
$M_{HI} = 2.35 \times 10^5 F D^2$, where $F$ is the flux integral in Jy\,\kms\ (\ie\ the integral $\int S\,dv$ where $S$ is the detected flux in a channel and $v$ is velocity) and $D$ is the distance to MO in Mpc.

The mass of \HI\ is approximately evenly divided between the two main blobs
straddling the jet (Fig.~\ref{fig:radhi}). The \HI\ column density
through the center of each of the blobs is $n_{HI} \sim 4.3 \times
10^{20}$\,cm$^{-2}$ and $n_{HI} \sim 4.7 \times 10^{20}$\,cm$^{-2}$ for
the northern and southern blobs respectively.  The total (projected)
extent of the \HI\ emission is approximately 275\,kpc$^2$.
 
\subsection{Keck LRIS spectroscopy}

Any potential spatial variations in the stellar populations or ages
are best studied through spatially-resolved spectroscopy. To
investigate any spatial variations in ionization and anomolous
kinematics across MO we obtained a high signal-to-noise longslit
spectrum using the Low Resolution Imaging Spectrometer
\citep[LRIS;][]{lris} on Keck I on 2004 January 20. The D560 dichroic
was used with a slit width of 1.5\arcsec. The total exposure time was
1800\,s. On the blue side, a 300 line mm$^{-1}$ grism, blazed at 5000\,\AA\
was employed, giving 1.43\,\AA\,pixel$^{-1}$ and spectral resolution
10.7\,\AA. On the red side, a 400 line mm$^{-1}$ grating, blazed at
8500\,\AA\ was used, giving 1.86\,\AA\,pixel$^{-1}$ and spectral resolution
6.8\,\AA. The data were reduced in \iraf, with cosmic ray removal, sky
line subtraction, wavelength calibration and flux calibration
performed in the standard manner.  The blue- and red-side spectra were
combined using {\sc scombine} in IRAF, and the resultant pixel scale
was 1.41\,\AA\,pixel$^{-1}$, so the resulting spectra were smoothed with a 3
pixel boxcar filter to improve signal-to-noise. Spectra were extracted using
several apertures and were corrected for Galactic extinction using
$E(B-V) = 0.045$ and $R_V = 3.1$ \citep{galred}.
\clearpage
\begin{figure*}
\centering
\includegraphics[width=0.45\linewidth,draft=false]{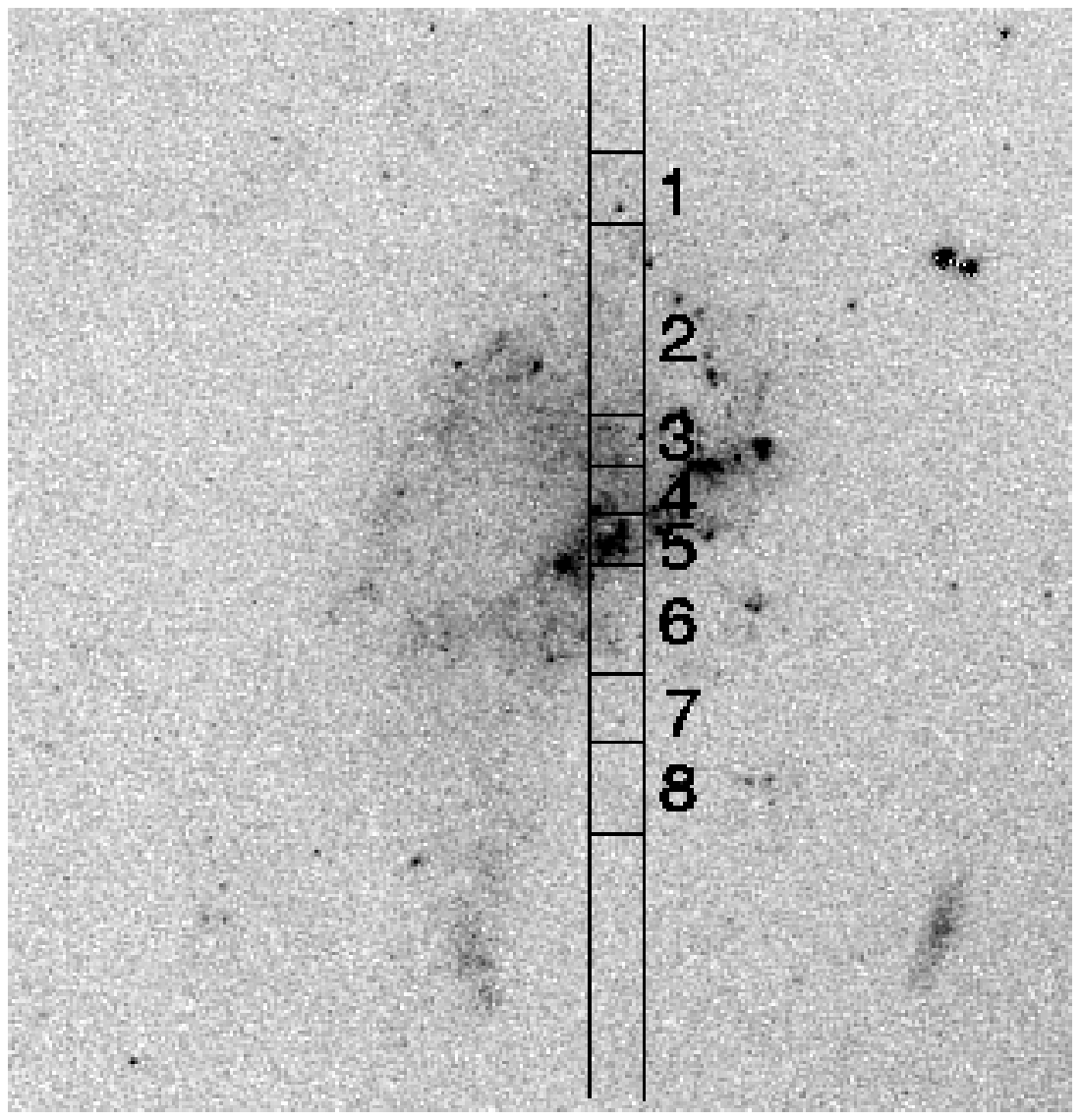}\hspace{0.05\linewidth}%
\includegraphics[width=0.45\linewidth,draft=false]{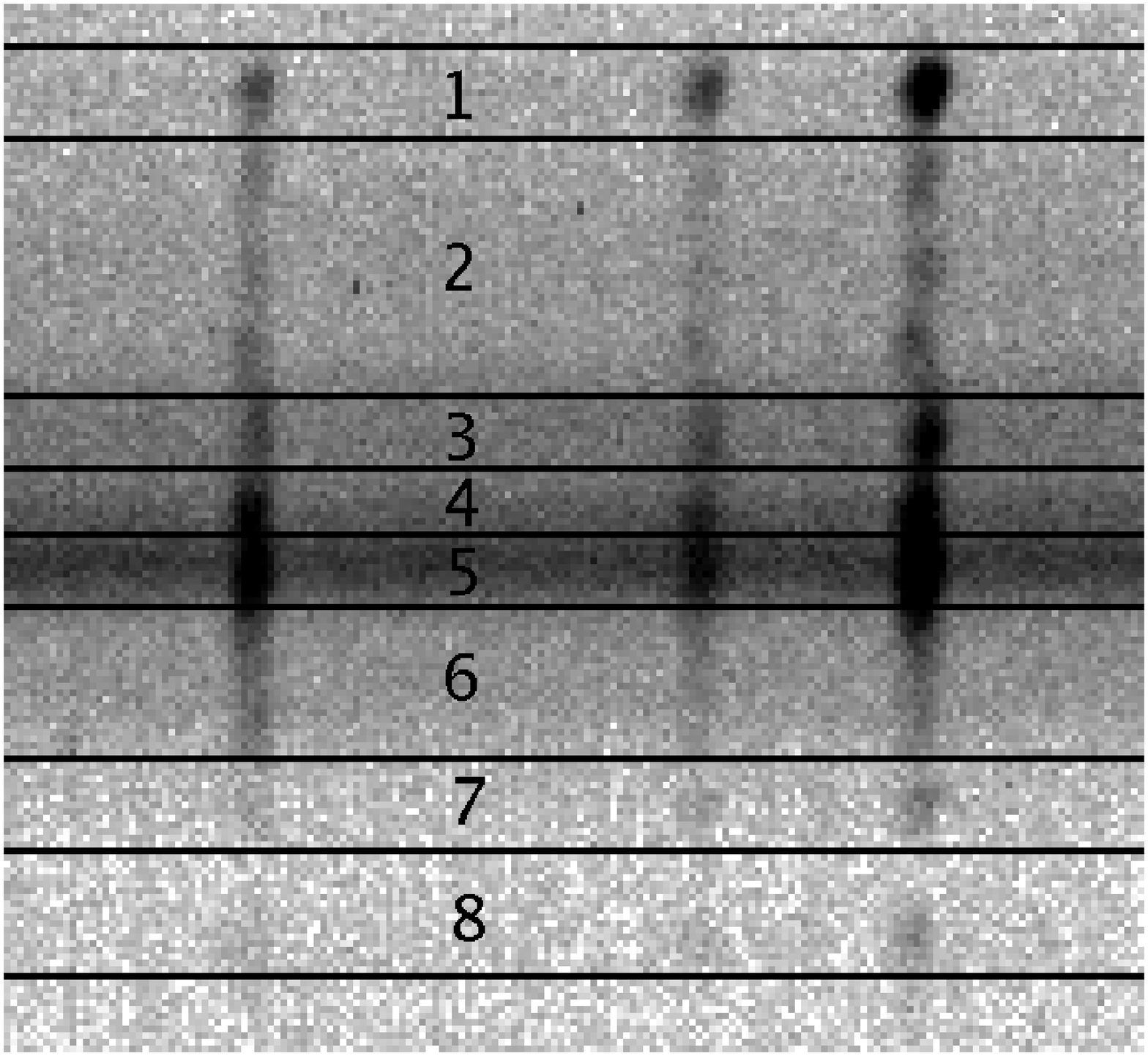}
\caption{\label{fig:lsq}The LRIS slit position ({\em left}), overlaid on the F555W image of MO (the width of the slit is accurately shown). Numbered sub-apertures are the same as shown on the section of the 2-D spectrum (showing \hb\ and the [\ion{O}{3}] doublet) at {\em right}, and the fluxes in table~\ref{tab:lines}. The ``full slit'' spectrum was extracted in an aperture covering subapertures 1 -- 8.
}
\end{figure*}
\clearpage
A large, 21\arcsec-long aperture spectrum (just covering subapertures 1 -- 8 as shown in figure~\ref{fig:lsq}) was extracted to obtain an
integrated spectrum encompassing all of the emission from MO along the
1.5\arcsec-wide slit. The extracted spectrum is shown in figure
\ref{fig:lris}. From this, the redshift of MO was determined to be $z
= 0.0189$ (similar to the value found by \citealt{wvb:85}). The
spectra were Doppler corrected to zero redshift. The measured emission line properties
integrated along the slit, and for various sub-apertures described
below are presented in Table~\ref{tab:lines}.

Fluxes were measured from Gaussian fits to the lines, with 1-sigma errors estimated from Monte Carlo noise simulations (noise estimates and gain were determined from measurements of the spectrum). 

The strengths of the
Balmer absorption lines (Table~\ref{tab:abslines}) were measured by
fitting a Gaussian to the wings of the absorption underlying the
Balmer emission components. Quoted fluxes are the amount of absorbed flux based on a fit to the continuum after subtraction of a Gaussian model for the emission component. The emission line fluxes in Table~\ref{tab:lines} are corrected for underlying absorption in the line cores.

\clearpage
\begin{deluxetable}{rrrrrrrrrrrr}
\tablecolumns{12}
\tablewidth{0pt}
\rotate
%\tablewidth{0pt}
%\tablefontsize{\scriptsize}
%\tabletypesize{\tiny}
\tablecaption{\label{tab:lines}Equivalent widths and Galactic-reddening-corrected line fluxes from LRIS spectroscopy}
\tablehead{
\colhead{Line} & \colhead{$\lambda_{rest}$ (\AA)} &  \colhead{$EW_0$ (\AA)} & \multicolumn{3}{l}{Flux (H$\beta =100$):} & \colhead{} & \colhead{} & \colhead{} & \colhead{} & \colhead{} & \colhead{} \\
\colhead{} & \colhead{} & \colhead{Full slit} & \colhead{Full slit}  & \colhead{Ap 1} & \colhead{Ap 2} & \colhead{Ap 3} & \colhead{Ap 4} & \colhead{Ap 5} & \colhead{Ap 6} & \colhead{Ap 7} & \colhead{Ap 8}
}
\startdata
$\!$[\ion{O}{2}] & 3727 & $63.6 \pm 0.4$ & 359.0     & 221        & 406       & 449        & 305          & 454        & 362       & 289      & 240  \\
\,[\ion{Ne}{3}]  & 3869 &  \nodata       &  \nodata &   \nodata & \nodata &  40         & 28          & \nodata & 31          &  \nodata &  \nodata \\
\,H8             & 3889 &  $1.0 \pm 0.3$ &   7.9        & \nodata & \nodata & \nodata  &   \nodata & \nodata & \nodata & \nodata & \nodata \\
\,H$\delta$      & 4102 &  $1.1 \pm 0.3$ &   9.7        & \nodata & \nodata &  \nodata &  \nodata &  \nodata &  \nodata &  \nodata &  \nodata \\
\,H$\gamma$      & 4340 &  $4.7 \pm 0.3$ & 40.8        & 51          &   31        & 33          & 30            & 31           & 30          & \nodata & \nodata \\ 
\,H$\beta$       & 4861 & $12.4 \pm 0.3$ & 100.0     & 100        & 100       & 100        & 100         & 100         & 100        & 100       & 100        \\
\,[\ion{O}{3}]   & 4959 &  $6.7 \pm 0.3$ & 60.7        & 115        & 115      & 59           & 85           & 49           & 36           & 67     & 87 \\ 
\,[\ion{O}{3}]   & 5007 & $19.4 \pm 0.3$ &172.4       & 333        & 333     & 180         & 252         & 169         & 114        & 211   & 200 \\ 
\,\ion{He}{1}    & 5876 &  $1.7 \pm 0.4$ & 9.4          & \nodata & 11        & 17            & 16           & 11           & 7            & \nodata &  \nodata \\ 
\,[\ion{O}{1}]   & 6300 &  $2.5 \pm 0.3$ & 14.1        & \nodata & 22        & 28            & 20           & 11           & 16          &  \nodata & 46 \\ 
\,[\ion{O}{1}]   & 6363 &  $1.0 \pm 0.3$ &  5.2         & \nodata & 7           & \nodata  & 7              & \nodata & \nodata & \nodata & \nodata \\ 
\,[\ion{N}{2}]   & 6548 &  $2.9 \pm 0.3$ & 13.6        & \nodata & 15         & 7             & 23           & 16           & 10         & \nodata & \nodata \\ 
\,\ha            & 6563 & $58.6 \pm 0.3$ & 278.7      & 226        & 306      & 335         & 396         & 318        & 238       & 350     & 390  \\ 
\,[\ion{N}{2}]   & 6583 &  $8.3 \pm 0.3$ & 39.5         & 14          & 44        & 54            & 59           & 46          & 36         & \nodata & \nodata \\ 
\,[\ion{S}{2}]   & 6716 & $17.1 \pm 0.3$ & 78.7         & 31          & 94         & 105         & 114         & 81           & 74         & 100        &  120 \\ 
\,[\ion{S}{2}]   & 6731 & $12.2 \pm 0.3$ & 55.0         & 18          & 55         & 75           & 73           & 57           & 53         & 50        &  \nodata \\ 
\,$F_{\rm {H} \beta}^{1}$ & & & $789 \pm 2$ & $84 \pm 3$ & $124 \pm 3$ & $57 \pm 3$ & $116 \pm 3$ & $140 \pm 3$ & $134 \pm 3$ & $ 18 \pm 3$ & $15 \pm 4$\\
\enddata
\tablecomments{$^{1}$ \hb\ line fluxes are given in 10$^{-18}$ \esc.}
\end{deluxetable}

\begin{deluxetable}{llll}
\tablecolumns{4}
%\rotate
%\tablewidth{0pt}
%\tablefontsize{\scriptsize}
%\tabletypesize{\tiny}
\tablecaption{\label{tab:abslines}Absorption lines from LRIS spectroscopy, corrected for Galactic reddening}
\tablehead{
\colhead{Line} & \colhead{$\lambda_{rest}$ (\AA)} & \colhead{Flux$^1$} & \colhead{$EW_0$ (\AA)}
}
\startdata
H10            & 3798 &  $  218 \pm 2$ & $  3    \pm 1    $ \nl 
\,H9             & 3835 &  $  262 \pm 2$ & $  3    \pm 1    $ \nl 
\,H8             & 3889 &  $  223 \pm 2$ & $  3    \pm 1    $ \nl 
\,H$\delta$      & 4102 &  $  260 \pm 2$ & $  3    \pm 1    $ \nl 
\,H$\gamma$      & 4340 &  $  363 \pm 2$ & $  4    \pm 1    $ \nl 
\,H$\beta$       & 4861 &  $  247 \pm 1$ & $  3    \pm 1    $ \nl 
\enddata
\tablecomments{$^{1}$Line fluxes are given in 10$^{-18}$ \esc.}
\end{deluxetable}
\clearpage

Additionally, spectra were extracted in eight subapertures, chosen to coincide with regions of the 2-D spectrum and images which appeared to have relatively homogeneous properties across them (see Fig.~\ref{fig:lsq}). The properties of the measured emission lines are also shown in Table~\ref{tab:lines}. There was insufficient signal-to-noise in the absorption lines in the subapertures to accurately determine absorption line strengths at these positions.
\clearpage
\begin{figure}
\centering
\includegraphics[width=\linewidth]{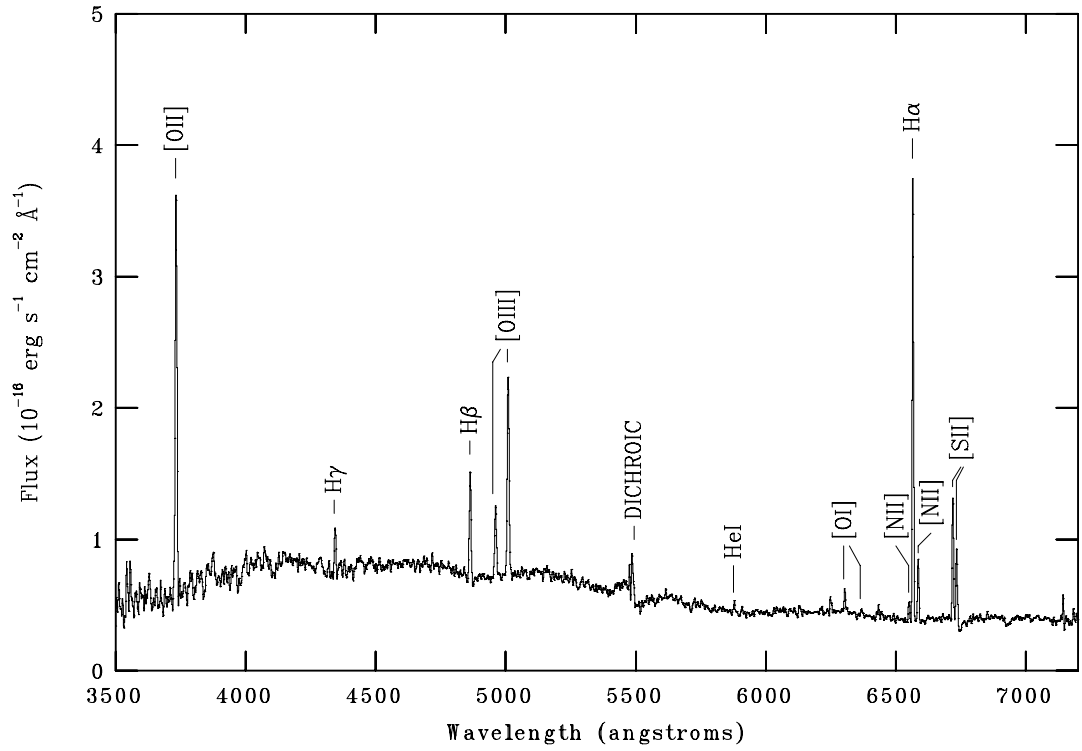}
\caption{\label{fig:lris} The combined blue and red side LRIS spectra
(1800\,s), smoothed with a 3-pixel boxcar filter.  }
\end{figure}
\clearpage

Line ratios which depend upon the measurement of lines on both the red
and blue sides are less reliable for the subapertures than for the
integrated spectrophotometry, due to the difficulty of measuring
exactly the same region of the 2-D spectrum in apertures of only a few
pixel length. This is presumably part of the source of the variation in the
estimated $E(B-V)$ seen in Table~\ref{tab:subapquant}, below. Line
ratios determined using lines from only the red or blue sides are
unaffected by this additional source of uncertainty.\label{sec:ebvsub}

Since the slit is oriented north -- south, the observations will to some extent be affected by atmospheric differential refraction \citep{difrefrac}. This was minimised by observing at relatively low airmass ($Z < 1.20$), but dispersion still has the effect of offsetting the physical regions of the source where lines at the red and blue ends of the spectrum are measured (as badly as 0.9\arcsec\ in the case of [\ion{O}{2}] and [\ion{S}{2}]). In the case of ratios such as \ha\,/\,\hb, the differential refraction is at worst around 0.36\arcsec, \ie, around 25\,\%\ of the width of the slit. Inspection of figure~\ref{fig:blobs} suggests that the extent of the extended emission compared to the slit width is such that a shift of this magnitude ought not to produce wild fluctuations in line ratios, especially since the seeing at the time of the LRIS observations (0.84\arcsec, some 5 times worse than that of the F555W image in figure~\ref{fig:blobs}) will tend to wash out features at smaller scales.

In an extreme case where the flux of \ha\ or \hb\ was different by a factor of two in the 25\,\%\ of the slit discussed above, this could lead to errors in the computed extinction values of $E(B-V)_s \sim 0.08$ (see section~\ref{sec:red}). The fact that the calculated value of $E(B-V)_s$ for the whole slit is so close to zero (formally, $E(B-V)_s = -0.01$) rather than, for example, strongly negative, suggests that these assertions are reasonable. This effect is presumably somewhat more important on smaller scales (\eg, the individual slitlets) and so may be an additional source of uncertainty (along with the difficulty of measuring the same region on the red and blue sides of the spectrum, discussed above) in the $E(B-V)$ values in table~\ref{tab:subapquant}. Since in practice, for the whole slit, we do not expect variations in flux of a factor two on scales smaller than the atmospheric seeing, we assume that atmospheric dispersion is not a major source of error in our analysis.

The astute reader will note that the total \hb\ flux quoted for the ``full slit'' in table~\ref{tab:lines} is around 15\% larger than the sum of the fluxes from the individual subapertures. This is again due to the difficulty of placing individual subapertures to sub-pixel precision, but as noted above does not affect ratios as measured in individual apertures.

The line ratios observed in these spectra confirm that the line
emission is due to \HII\ regions throughout MO \citep{vo:87,dopita:00}. In particular, there
is no evidence of shock-excited gas, which would reveal itself by very
large [\ion{N}{2}]\,/\,H$\beta$ or [\ion{O}{2}]\,/\,H$\beta$ ratios \citep{dopita:84}. Nor does
the [\ion{O}{1}]\,/\,H$\alpha$ ratio become sufficiently large to put it
in the shock-excited classification \citep{ds:95}. Furthermore, the kinematics show
that the ionized gas is quiescent. There is no evidence for large
velocity gradients, at least across the jet in the north--south slit
direction. However, since shock-excited gas is more likely to reveal
itself at low flux levels, it might be of interest to obtain high
signal-to-noise spectroscopy along the jet direction.

\section{Derived Quantities}

\subsection{Reddening}

\label{sec:red}
The measured Balmer line ratio \ha\,/\,\hb\ from the full LRIS spectrum
(after correction for Galactic reddening), was found to be $2.790 \pm
0.008$. Following \citet{cal:94} and \citet{cal:01}, this is consistent with an intrinsic (continuum) reddening of $E(B-V)_s \sim 0$.

We also calculated values of $E(B-V)_s$ for the subapertures (see table~\ref{tab:subapquant}). These show some variation, including values as negative as $-0.10$. We believe these determinations to be less reliable than the full spectrum value for reasons discussed in section~\ref{sec:ebvsub}; errors introduced by atmospheric dispersion and by not measuring the same part of the spectrum in the red and blue sides can probably account for any deviations from $E(B-V)_s = 0$ in the subapertures.

We likewise used the (Galactic reddening corrected) broad-band UV
fluxes from GALEX to calculate $\beta$, the UV continuum slope
\citep{cal:94}, and obtained a value of $\beta = -1.734$, again
yielding $E(B-V)_s \sim 0$, in excellent agreement with the line ratio
results.

These more accurate measurements imply that there is essentially 
no dust obscuration along the line of sight to the emission line regions of MO and supersede those of \citet{wvb:85} who obtained $E(B-V) = 0.27$. It is possible, however, that some of the variation in calculated values of $E(B-V)$ in the subapertures is in fact due to the presence of localised dust. If the dust is mixed in with the emission line gas and stars to some extent (rather than a foreground screen, which is the assumption in the extinction calculations) then dust mass can be somewhat higher for a given $E(B-V)$ than otherwise.

\subsection{Metallicity}

As remarked above, the emission-line ratios are typical for those of
\HII\ regions, showing that the emission is dominated by star forming
regions and so confirming the earlier conclusions by
\citet{wvb:85}. The high quality of our data allows us to obtain
additional, more accurate diagnostics.

We determined the metallicity of MO using the method outlined in
\citet{kewley:02}. This yields a value of $log(O/H) + 12 = 8.6 \pm
0.1$, equivalent to a metallicity $Z \sim 0.5\,Z_{\odot}$. The mean
ionisation parameter was determined to be $q \approx 3 \times
10^7$\,\cms\ ($\ionpar = 10^{-3}$). This is within the normal values
expected for extragalactic \HII\ regions \citep{dopita:00}.

\subsection{Star Formation Rates}

Applying the relation between \ha\ luminosity and star formation rate
(SFR) \citep{dopita:94,kennicutt:98,panuzzo:03},

\begin{displaymath}
\left(SFR_{{\rm H\alpha }} \over \rm {M_{\odot}~yr^{-1}}\right)
=(7.0-7.9)\times 10^{-42}\left( L_{{\rm H\alpha }}\over {\rm erg\,s}^{-1}\right),  \label{1}
\end{displaymath}

we used the measured \ha\ flux from the ATT images to obtain a total
SFR within MO of $\sim 0.5$\,\msunpyr. This estimate is sensitive to
the form of the initial mass function assumed for high mass stars, and
therefore has a uncertainty of order 30\%.

The local star formation rate increases, as expected, in the
subapertures near the central blobs, where it is up to an order of
magnitude higher than in the subapertures in the outer regions of MO
. The measured star formation rates are given in
Table~\ref{tab:subapquant}, assuming the globally determined result of no intrinsic dust.
\clearpage
\begin{deluxetable}{lllllllll}
\tablecolumns{9}
%\rotate
%\tablewidth{0pt}
%\tablefontsize{\scriptsize}
%\tabletypesize{\tiny}
\tablecaption{\label{tab:subapquant}Derived quantities from the observed emission lines from the LRIS subapertures}
\tablehead{
\colhead{Quantity} & \colhead{Ap 1 } & \colhead{Ap 2 } & \colhead{Ap 3 } & \colhead{Ap 4 } & \colhead{Ap 5 } & \colhead{Ap 6 } & \colhead{Ap 7 } & \colhead{Ap 8}
}
\startdata
$E(B-V)$ & -0.10 & 0.03 & 0.06 & 0.13 & 0.04 & -0.08 & 0.08 & 0.13 \nl
Subaperture width (\arcsec) & 1.89 & 5.13 & 1.35 & 1.22 & 1.35 & 2.97 & 1.78 & 2.43 \nl
Subaperture size (kpc$^2$) & 0.397 & 1.076 & 0.283 & 0.256 & 0.283 & 0.623 & 0.373 & 0.510 \nl
SFR (\ha) $10^{-3}$ \msunpyr\,kpc$^{-2}$ & 2.90 & 2.14 & 4.09 & 10.90 & 9.56 & 3.12 & 1.02 & 0.70 \nl
\,[\ion{S}{2}] 6716 / 6731 & 1.73 & 1.71 & 1.40 & 1.56 & 1.43 & 1.39 & 2.00 & \nodata \nl 
log $\ionpar$ \tablenotemark{a} & -2.6 & -3.0 & -3.2 & -3.0 & -3.0 & -3.0 & \nodata & \nodata \nl
\enddata
\tablenotetext{a}{Determined from the [\ion{N}{2}] / \ha\ diagnostic \citep{kewley:02}, assuming constant metallicity of $Z = 0.5 Z_{\odot}$}
\end{deluxetable}
\clearpage
\subsection{Electron Densities}

The density-sensitive [\ion{S}{2}] $\lambda 6716 / \lambda 6731$
doublet ratio was measured to be $1.431 \pm 0.01$ in the full spectrum. No other electron
density diagnostics were able to be determined from our LRIS spectra,
due to insufficient spectroscopic resolution and / or
sensitivity. Using the five-level atom model of \citet{derob:87}, and
assuming an electron temperature, $T_e = 10^4$\,K, this yields an
electron density of order 1 -- 10\,cm$^{-3}$, averaged along the
slit. However, for electron densities lower than approximately
100\,cm$^{-3}$, corresponding to values of this ratio $\gtrsim 1.32$,
this line ratio becomes rather insensitive to electron density, and
our measured electron density is therefore highly
uncertain. Measurements of the ratio in the individual subapertures
yield values ranging from 1.39 -- 2.00
(Table~\ref{tab:subapquant}). Given that the errors on the ratios are
rather high, due to low signal-to-noise in some subapertures, and that
the \ion{S}{2} lines are almost coincident with strong telluric lines
whose subtraction introduces additional uncertainties, all we can
conclude with certainty is that the electron densities are less than
100\,cm$^{-3}$.

\subsection{Ages of the Clusters}
\label{sec:ages}

As \citet{gd:99} have shown, the strengths of the \ion{He}{1} lines
relative to H$\beta$ are sensitive to the age of an \HII\ region. We created
our own photoionization models (appropriate to the metallicity of MO
determined above) using {\sc mappings iii}r \citep{sd:93,dopita:02} in combination with \emph{Starburst99} \citep{sb99} stellar
synthesis instantaneous burst models, with a Salpeter IMF with
lower mass limit of 1\,\msun\ and upper mass limit of
100\,\msun. The use of a Salpeter IMF is appropriate for the modelling of both
the ionizing flux and stellar continuum for a young ($<10^8$\,yr) recent
starburst. The neglect of the $<1$\,\msun\ stars makes no difference to the
results, as they contribute negligible luminosity at these
early times. We find that the \ion{He}{1} $\lambda 5876$ /
H$\beta$ line ratio remains near 0.1--0.12 until 4\,Myr, after which it
declines steeply, becoming $<0.05$ after 5\,Myr. The measured line
ratios in table~\ref{tab:lines} are therefore consistent with \HII\
region ages of $<5$\,Myr, for all sub-apertures.

The [\ion{O}{3}]$\lambda 5007$ / H$\beta$ ratio is even more
sensitive to age. In our models, it starts at around 4.5, declines to
$\sim 1.2$ by 2.5\,Myr, to 0.5 by 3.0\,Myr, and is negligible
thereafter. We can therefore conclude that all the observed \HII\
regions are younger than $\sim 3$\,Myr.

Comparison of the equivalent widths (EWs) of the Balmer nebular emission
lines (table~\ref{tab:lines}) with figure~16 of \citet{gd:99} rules out continuous starburst
models. The measured \hb\ and H8 EWs both yield an
instantaneous burst age of 6\,Myr, and H$\delta$ yields an
instantaneous burst age of 8\,Myr.  Comparison of the measured
EWs with figures 3 and 4 of \citet{gd:99} shows that
ages greater than around 10\,Myr are excluded.

The discrepancies between the ages obtained from [\ion{O}{3}]$\lambda 5007$ / H$\beta$ and those from the EWs (discussed above) and SED fitting (Section~\ref{sec:sed}) may be due to a combination of some ongoing star formation, or possibly some contribution to the [\ion{O}{3}] luminosity from shocked gas. Spectra with higher spatial resolution and signal-to-noise might help resolve this.

\subsection{Number of Ionizing Photons}

Taking the Galactic reddening corrected ratio \ha\ / \hb\ = 2.79
determined from the full-aperture spectroscopy, and the \ha\ flux for the whole
object from the imaging, we derive an \hb\ luminosity of $2.37 \times
10^{40}$ \es. Comparing to \citet{gd:99} table 7, 
and assuming an age of 6\,Myr (from the emission line EWs), we obtain a
stellar mass of $1.9 \times 10^7$\,\msun, and log $Q_*$ (the number
of Lyman continuum photons) of 52.7. If we were instead to use the age of 3\,Myr obtained from the \HII\ region spectroscopy, this would yield a stellar mass of $1.5 \times 10^6$\,\msun, so this determination is quite sensitive to age. The value of log $Q_*$ depends only on the \hb\ luminosity and so is independent of age.

From \citet{sl:96},

\begin{displaymath}
\ionpar = 2.8 \times 10^{-20}(10^4/T_e)^{2/3}(Q_*n\epsilon^2)^{1/3}
\end{displaymath}

and with the measured values $\ionpar = 0.001$, $n = 10$\,cm$^{-3}$,
log $Q = 52.7$, and assuming $T_e = 10^4$\,K, we obtain a nebular gas
filling factor $\epsilon \sim 0.01$.

\subsection{Photometric Properties}
\label{sec:photom}

The NUV, FUV, ESI-R, F814W, $J$ and \kp\ images were registered to the
F555W image using the individual astrometric solutions. The F555W,
F814W, $J$ and \kp\ images were convolved to match the 0.72\arcsec\
seeing of the ESI-R image. We refrained from convolving all images to
match the 5.6\arcsec\ FWHM of the NUV image since this would result in
such a significant loss of spatial resolution for the majority of the
images. Photometry was performed in an 8\arcsec\ radius circular
aperture to enclose all the emission from MO. Since MO is resolved in
the GALEX images (Section~\ref{sec:galex}) we also obtained photometry for three sub-apertures (Fig.~\ref{fig:threephot}) centered on MO-North, the NE filament, and MO-South (Section~\ref{sec:timing}).

The elliptical galaxies in the ESI image (Fig.~\ref{fig:esipp}) were
modelled using the tasks {\sc ellipse} and {\sc bmodel} in \iraf, and
the best fitting models were subtracted from the original
image. NGC\,541 is difficult to fit in this image because it extends
beyond the boundary of the frame to the right, but we were able to do
a reasonable job of subtraction by carefully choosing and refining the
fit parameters (see Fig.~\ref{fig:esipp}).

\begin{figure}
\centering
\includegraphics[width=\linewidth]{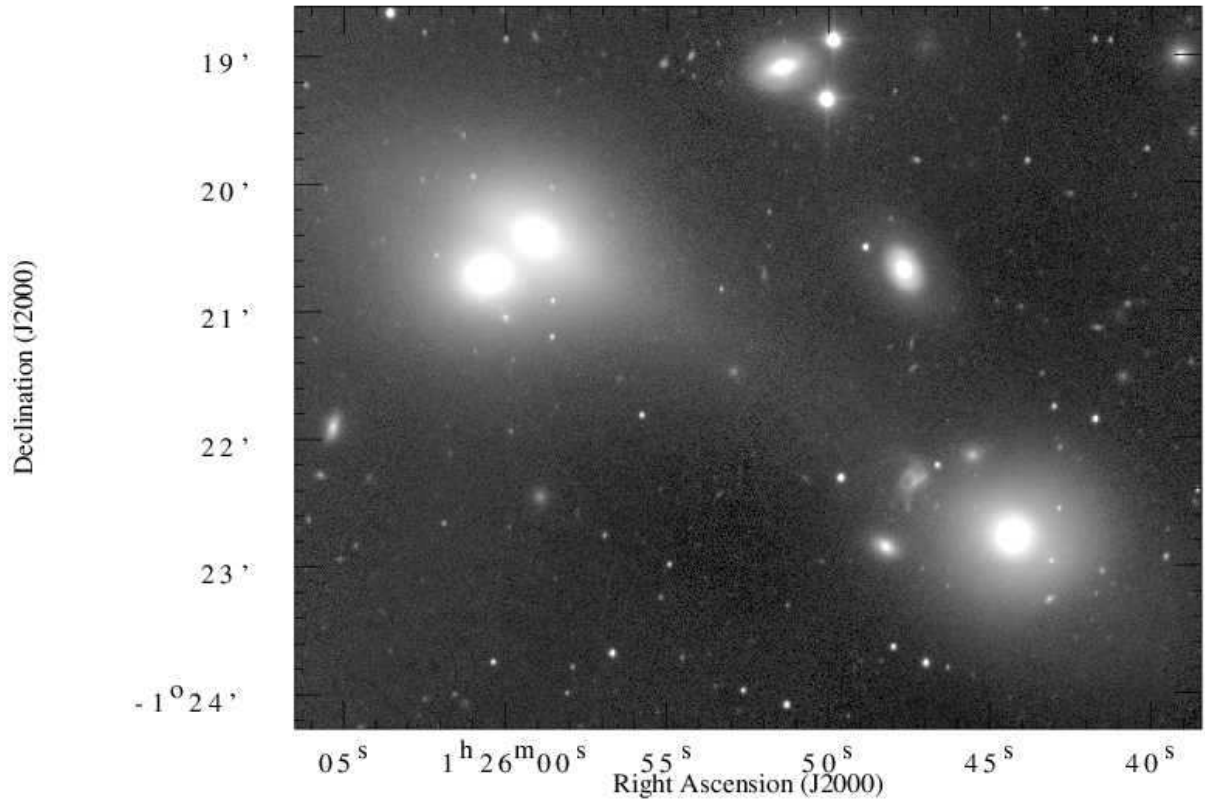}
\caption{\label{fig:pfcamr} PFCam $R$-band coadd, cropped to highlight
the faint bridge of stars which connects NGC\,541 with the interacting
pair NGC\,545 / 547. MO lies in this bridge, close to NGC\,541.  }
\end{figure}

After subtraction of the elliptical galaxies, the bridge of stars
joining NGC\,541 and NGC\,545 / 547 (see also Fig.~\ref{fig:pfcamr})
can be more clearly seen and more accurately measured. The bridge is
around 24.5 mag arcsec$^{-1}$, and extends over around 115\arcsec
$\times$ 45\arcsec\ ($\sim 700$\,kpc$^2$); its total R-band luminosity
is around $3 \times 10^9$\,\lsun. From \citet{a194}, the gas mass
within a 0.8\,Mpc radius of the center of the cluster is $1 \times
10^{13}$\,\msun, so we expect (by simple scaling, in the lack of more detailed
information about the gas distribution) the gas mass in the bridge to be
approximately $3 \times 10^9$\,\msun.

\clearpage
\clearpage

The bridge is approximately 18.7\,mag in the aperture used for the
broadband photometry (\ie, approximately 15\%\ as bright as MO) and
the portion of the NGC\,541 model enclosed by this aperture is $55^{+23}_{-20}$~\%\ as bright as MO. Hence
about 40\,\%\ of the light in the aperture comes from NGC\,541 and the
bridge. Since a circular aperture was used, this includes parts of the
aperture where MO is relatively faint, and parts where it is
relatively bright, and so cannot be construed to be typical for the
spectroscopic apertures or the individual pixels of the
image. However, we consider the contribution from the bridge and
elliptical when performing spectral synthesis
modeling.\label{sec:galmod}

\subsection{Spectral Energy Distributions}
\label{sec:sed}

We computed SEDs using the seven bands of integrated (8\arcsec\ radius
aperture) Galactic reddening corrected photometry for MO, converting
magnitudes to fluxes in \esca\ using published zeropoints, and
normalising the fluxes to the flux measured in the F555W filter
(approximately $V$-band) to facilitate model comparisons. A set of
model spectra were created using \emph{Starburst99}, with parameters as
described in section~\ref{sec:ages} . These models were redshifted to match
the redshift of MO,
and convolved with the filter response functions, resulting in a set
of fluxes for each model. These were normalised to the F555W flux, and
the best match between these fluxes and those from the observed
photometry were determined by a chi-squared minimisation procedure. 

\begin{figure*}
\centering
\includegraphics[width=2.9in]{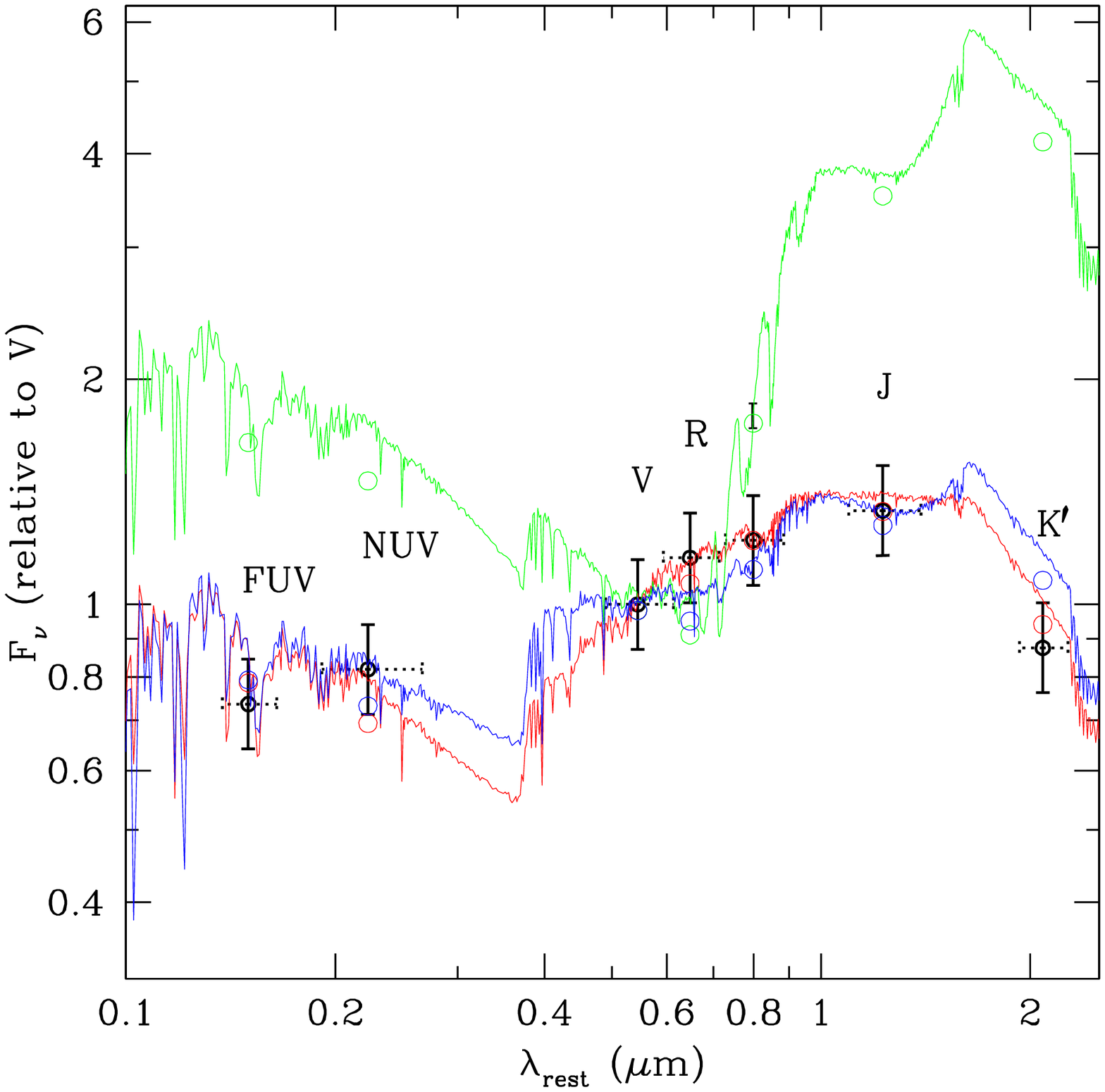}\hspace{0.5in}
\includegraphics[width=2.9in]{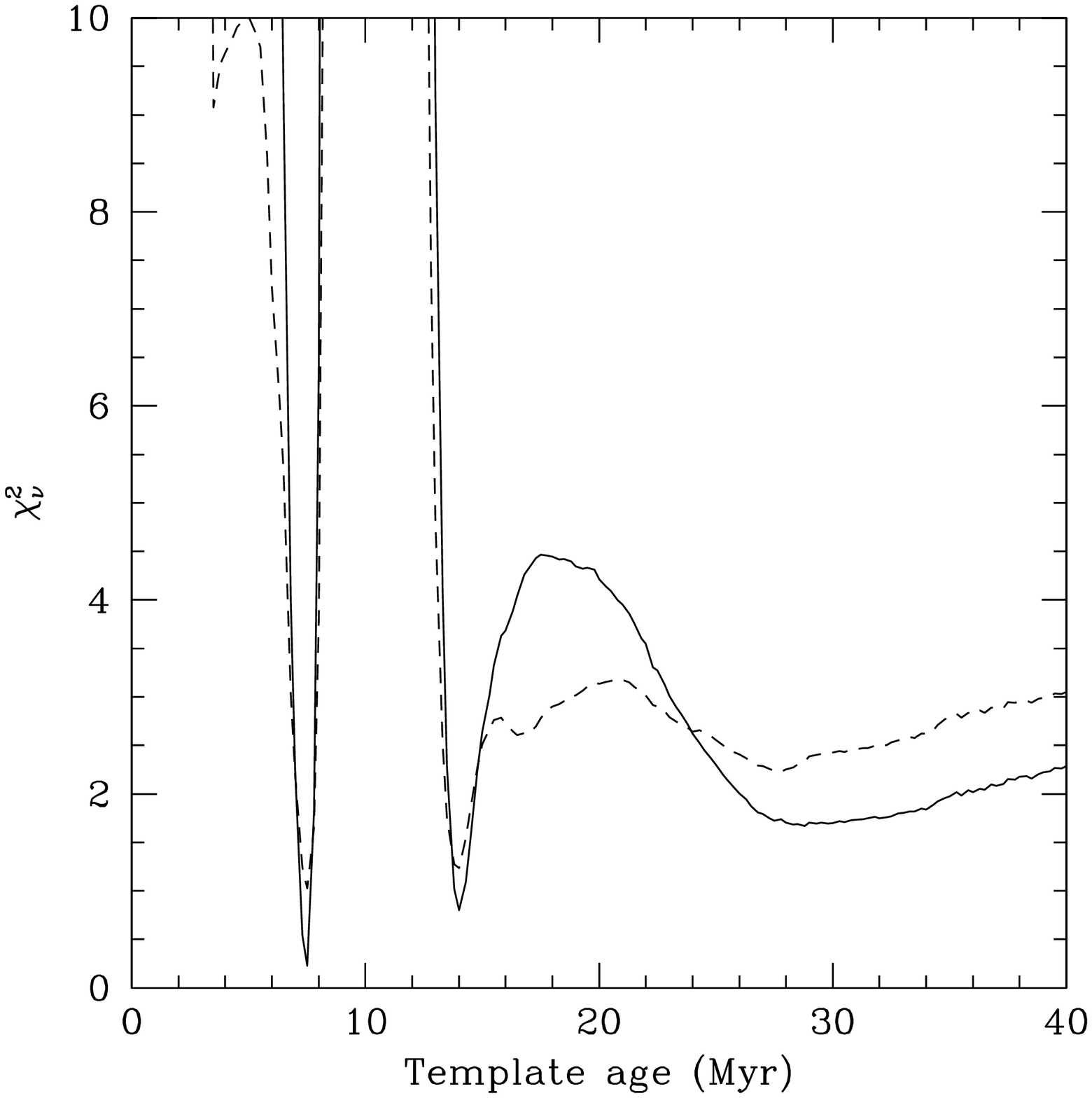}
\caption{\label{fig:mosed} {\em Left:} A selection of synthetic
spectrum models (solid lines) and the photometry computed by
convolving those models with the filter transmission curves (colored
open circles), is compared to the observed (Galactic reddening
corrected) photometry (black points). The vertical errorbars on the
black points represent measured photometric errors, and the horizontal
bars represent the filter effective bandwidths. The best fitting
(7.5\,My) model is shown in red, along with a badly fitting model
(9.5\,My in green) and another model which provides quite a good fit,
\ie\ low chi-squared (14.0\,My in blue). {\em Right:} Reduced chi-squared for the model fits, as a function of time since the
instantaneous burst in the spectral template (solid line). There is a clear minimum
around 7.5\,My, and another broad, but less deep, minimum for ages
older than 14\,My. The old models fit the photometry at $V$ through
\kp\ quite well, but do a poor job of reproducing the UV flux. Adding a 1\,Gyr-old population at the 40\,\%\ level produces \chisq\ values shown as the dotted curve.}
\end{figure*}

In figure~\ref{fig:mosed}, we plot $\rchisq = \chisq / \nu$, where $\nu = 6$ is the number of
degrees of freedom (since seven filters are used, and the normalisation takes away one degree of freedom). 
The best fitting (single-age) model was a 7.5\,Myr old SED with $\rchisq = 0.22$. Models older than 14\,Myr fit the photometry
at $V$ through \kp\ quite well, resulting in a broad minimum in the
\chisq\ fit ($2 \lesssim \rchisq \lesssim 4$). However, these older models do a poor job of
reproducing the UV flux, and hence have higher \chisq\ than the 7.5\,Myr model, which results in a sharp minimum
in \chisq, and reproduces the UV fluxes much better. A $\Delta \chisq$ analysis, following \citet{avni:76} (with the number of ``interesting parameters'' equal to one, \ie, the template age), yields formal 68\,\%\ confidence 
limits of $7.47 \pm 0.12$\,Myr (interpolating \rchisq\ values between the 0.25\,Myr template age spacing). The secondary minimum at 14.02\,Myr ($\rchisq = 0.80$) is statistically a worse fit at $> 90$\,\%\ confidence. The associated \chisq\ probability for the primary minimum is 0.97.

The two disjoint regions of low chi-squared are due to the similarity of the SEDs at 7.5 and 14.0\,Myr. Between these ages, the UV and infrared fluxes increase above the model values due to the dominance of red supergiants in the models. As mentioned above, older models ($\gtrsim 25$\,Myr) fit the optical and infrared points reasonably well, but do not do a good job of fitting the UV points. 

In an attempt to account for an underlying (presumably old) population
from the bridge and NGC\,541 (see section~\ref{sec:galmod}) we ran the
\chisq\ minimisation procedure again, this time adding in a 1\,Gyr
population at the 40\,\%\ level (the number of degrees of freedom is unchanged since the
normalisation of this component is fixed by the photometry from section~\ref{sec:galmod}). The best fit for the remaining
component was still the 7.5\,Myr model, with $\rchisq = 1.02$ (slightly
higher than for the single population model, but with a \chisq\ probability 0.41). This yields 68\,\%\ confidence limits of $7.50^{+0.17}_{-0.18}$\,Myr, with a secondary minimum at 13.94\,Myr ($\rchisq = 1.23$) statistically distinct from the primary solution only at $> 68$\,\%\ confidence. 

The old population fraction was increased in steps of 1\,\%, and the \chisq\
procedure run over again. The \chisq\ minimum continued to
increase (as the UV data points were less well fit), but the 7.5\,Myr
model was still selected. At an old population fraction above 50\,\%,
the best-fitting young population begins to fluctuate between models
and the fit becomes increasingly poor. It seems reasonable, therefore,
to conclude that less than 20\,\%\ (and possibly as low as 0\,\%) of
the light (although possibly a larger percentage of the stellar mass)
in the aperture is due to an old population in MO itself, as suggested by the photometry in section~\ref{sec:photom}.

\begin{figure*}
\centering
\includegraphics[width=0.55\linewidth]{f14a.eps}\hspace{0.05\linewidth}
\includegraphics[width=0.35\linewidth]{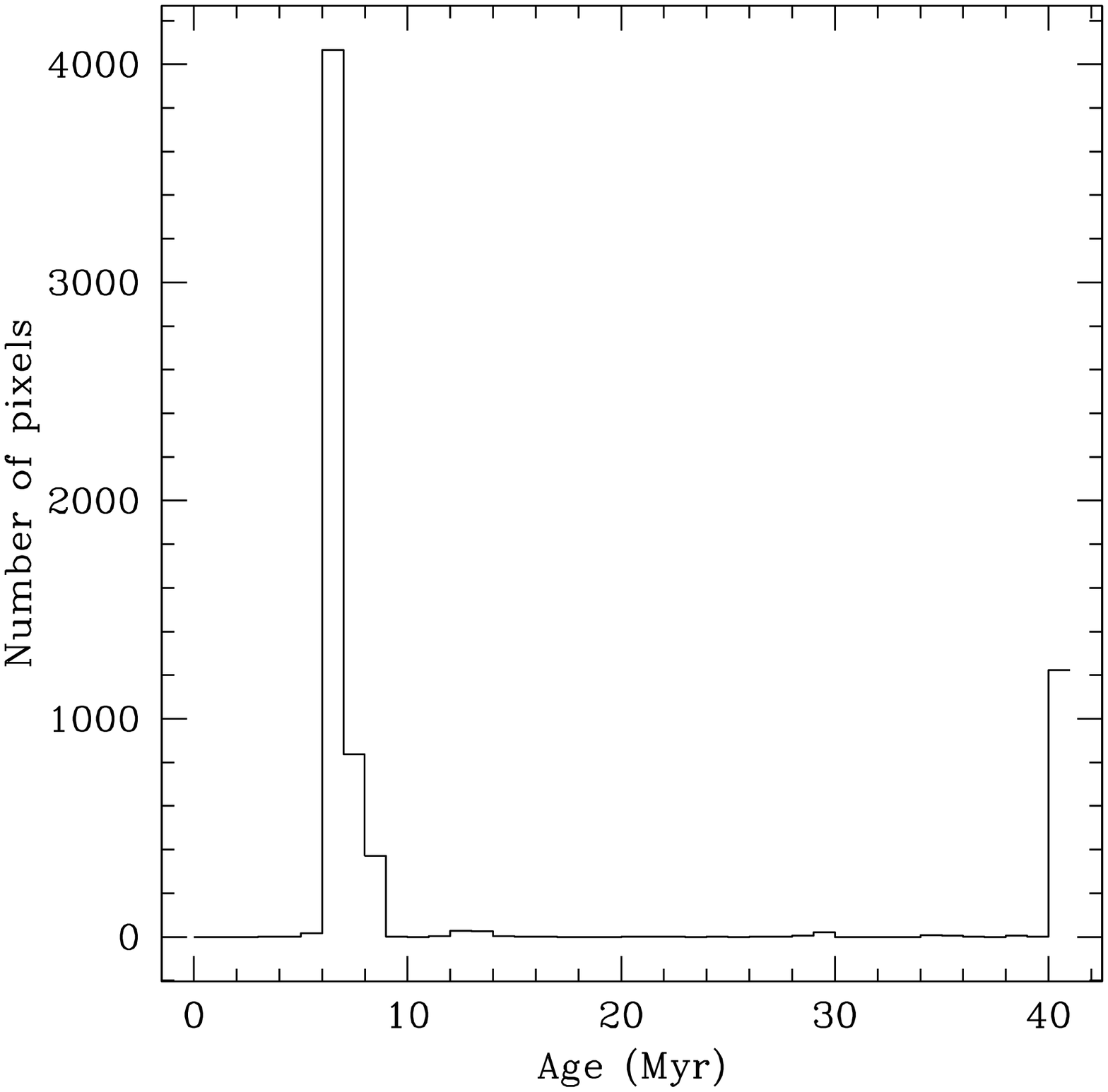}
\caption{\label{fig:pixage} {\em Left:} Pixel age map for MO. A range
in ages is seen in MO, color coded according to the wedge at right. Dark blue pixels were rejected due to their large \chisq\ (meaning they were poorly fit). Red pixels were best fit by models of $\geq 9$\,Myr (this includes ages up to and including 1\,Gyr). White contours are from the binned HST F555W image, and are overlaid to show the relationship between the morphology of MO and the clustering of pixels of the same age. Low signal to noise results
in some spurious pixels, but broadly it appears that pixels of the
same age cluster together, with the oldest populations in the central
bar of MO, and younger populations in the outer regions. This picture
is supported by binning pixels in these regions to improve
signal-to-noise (see text). {\em Right:} Histogram of the best fitting
pixel ages. Most of the older pixels are in the nearby old elliptical
NGC\,541. The peak at 40\,My includes the 1\,Gyr old pixels.}
\end{figure*}

We used a technique similar to that employed by \citet{johnston:05} to
obtain an age map for the galaxy. The registered, seeing-convolved
images were binned 4 by 4 pixels. The model fitting routine, as
described above, was run on the SEDs extracted from individual
corresponding binned pixels in each band, producing a best-fitting age
for each pixel (again assuming a single-age instantaneous
starburst). The resulting pixel age map is shown in
Fig.~\ref{fig:pixage}.
\clearpage

\clearpage
Reassuringly, pixels of the same age cluster together. There appears
to be a gradient in MO, from pixels with ages around 8.5\,Myr at the
center, to pixels with ages around 7.0\,Myr at the edges. The edge of
NGC\,541 at the lower right of the frame, and also the nearby small background galaxy, are fit with a 1\,Gyr old
population (the oldest population considered in our fits). We are
confident, therefore, that the pixel ages are representative of the
dominant population. Since young stellar populations tend to dominate
the integrated light of a galaxy, we cannot rule out an underlying old
population, and indeed a small number of pixels in MO itself are best fit with older populations. However, this may be simply due to low signal-to-noise in the UV data in these pixels, which provides less discrimination between the primary and secondary minima in the \chisq--age plot (the primary minimum tends to remain relatively sharply defined, whereas the secondary minimum becomes more broad, compared to the plots in figure~\ref{fig:mosed}). The pixel ages are consistent, however, with a picture where jet-induced star formation propagates outward from a region near the central ``bar'' in MO.

We also extracted SEDs from the registered, convolved, binned
multi-wavelength images in three boxes at the locations shown in
Fig.~\ref{fig:threephot}, corresponding to MO-South, MO-North, and the
filamentary structure to the north of MO. Following the same procedure
as for the pixel age map, but using these larger boxes to improve
signal-to-noise, we obtain ages (with 68\,\%\ confidence limits) of $6.66^{+0.44}_{-0.31}$\,Myr, $7.86^{+0.13}_{-0.21}$\,Myr, and  $7.00^{+0.46}_{-0.33}$\,Myr
respectively for the three components. These more sensitive
measurements support the conclusion from the pixel age map that
MO-North was the first region to form stars, and that star formation
spread over $\sim 1$\,Myr to the regions further out (MO South and the NE Filament have the same age to within these quoted errors). Presumably the
star formation will continue to advance ``downstream'' in future, as
the gas seen in the \hi\ map begins to form stars.

\section{Discussion}

Our new observations of MO reinforce the conclusion by \citet{wvb:85}
that MO is a young starburst triggered by the radio jet from
NGC\,541. However, some interesting questions remain -- firstly, what
existed at that location before the jet entered? Possibilities range
from a pre-existing gas-rich galaxy, a cold gas cloud, or else a warm
or even hot diffuse cloud of gas. Secondly, what are the physical
conditions in the jet (speed, Mach number, mass loading etc.) such
that it is able to trigger this starburst? Our observations add some
further insight into the first question. To learn more about the
second will probably first require a better understanding of FR-I jets
in general \citep{young:05,eilek:02,kataoka:06} and, ultimately,
numerical simulations of these types of jets.

\subsection{Kinematics and morphology}

The neutral hydrogen emission is intimately associated with
the radio jet: the \HI\ appears to be ``wrapped around" the axis of
the jet, and there is no \HI\ outside the jet! It will be important to
search for possible very low surface brightness \HI\ that might exist
near MO, in the bridge or elsewhere, since our observations were taken
in C-array and very large scale emission might have been
missed. But given the current observational evidence, there is good
reason to believe that there was no extended \HI\ prior to the jet
interaction, and therefore that the \HI\ is entirely the result of
radiative cooling of warmer gas in the inter-galactic medium triggered
by the shocks induced in it by the passage of the radio jet.

Numerical simulations of jet / cloud collisions, subject to
assumptions about the radio jet and the multi-phase medium with which
it interacts, show that radiative cooling may indeed lead to star
formation \citep{mo_sim}. The mass inferred for the \HI\ cloud is $4.9 \times 10^8$\,\msun, and the projected extent of the \HI\ emission is 275\,kpc$^2$ (section~\ref{sec:hi}).
The total amount of (hot) gas in this area according to the observations of \citet{a194} is expected to be $\sim 1.4 \times 10^9$\,\msun, so the available reservoir is certainly large enough. 
The question is: how
did it cool to form the \HI? Until we have better X-ray data we will
have to assume that the IGM in the Abell~194 cluster resembles to some
extent that of so-called cooling flow clusters, which contain both hot
($10^6 - 10^7$\,K) and cool / warm ($10^4$\,K) gas.  We note Abell~194
is reportedly a ``poor and cold'' cluster \citep{a194} and as such
a population of relatively cool, dense clouds in its IGM may well
exist.

The jet-\HI\ relationship in MO is different from that in Centaurus~A,
where the jet is running into a pre-existing, very extended \HI\
structure found in rotation around NGC\,5128 \citep{schim:94}. Another
difference is that in MO the \HI\ is downstream and the star formation
is closer to the AGN, while in Cen~A it is the other way around:
the \HI\ is closest to the AGN and the star formation is
downstream. Perhaps this is the difference between a jet hitting an
existing \HI\ cloud (Cen~A) and the jet causing the warm gas to
cool to form \HI\ (MO).

The \HI\ shows a significant velocity gradient (up to 100\,\kms) at
the jet / cloud interaction site \citep{cena}. Other examples, of even
larger \HI\ outflows (1000\,\kms) have recently been discovered in a
number of radio galaxies, to the extent that these may remove
significant amounts of gas from their parent galaxies
\citep{morganti:05}. However, these are all very luminous far-infrared
galaxies with lots of ISM cold gas and dust through which the jets are
propagating. In these cases it is reasonable to assume that the \HI\
gas may have been entrained in some way, as it may also have in
Cen~A \citep{cena}.

In MO the velocity gradient is much smaller, $\sim 40$\,\kms, but the
scenario may be qualitatively the same although in MO the entrained
gas clouds were probably not cold but warm. However, also in this case
the clouds will be accelerated, while they cool radiatively. Following
\citet{mo_sim} the velocity $v_{cl}$ of the accelerated clouds is

\begin{displaymath}
\left(\frac{v_{cl}}{\kms}\right) \lesssim 20\left(\frac{\chi}{10^3}\right)^{-1}\left(\frac{R_{cl}}{100\,{\rm pc}}\right)^{-1}\left(\frac{v_{sh,b}}{10^3\,\kms}\right)^2\left(\frac{t}{\rm Myr}\right)
\end{displaymath}

% The same equation, displayed in an alternate way:
%\begin{displaymath}
%\left(\frac{v_{cl}}{\kms}\right) \lesssim \frac{20\left(\frac{v_{sh,b}}{10^3\,\kms}\right)^2\left(\frac{t}{\rm Myr}\right)}{\Big(\frac{\chi}{10^3}\Big)\left(\frac{R_{cl}}{100\,{\rm pc}}\right)}
%\end{displaymath}

where $\chi = n_{cl}/n_{b,i}$ is the density contrast between the
cloud and background intercloud medium; $R_{cl}$ is the cloud radius;
$v_{sh,b}$ is the shock velocity in the background intercloud medium,
and $t$ is the time during which the cloud is being accelerated. This
time will be at most $t_{acc} \sim \sqrt{\chi} t_{cc}$ yr, where
$t_{cc}$ is the cloud crossing time for the radiative shocks. For the
parameters used by \citeauthor{mo_sim} one finds $t_{cc} \sim 1$\,Myr,
$t_{acc} < 30$\,Myr and $v_{cl} \sim 10 - 100$\,\kms. Thus within the
assumed framework of a jet interacting with a clumpy warm medium it is
possible to explain the observed \HI\ velocity shear of $\sim
40$\,\kms.

The VLA \HI, GALEX UV, and ATT \ha\ images all reveal that MO has a
double structure which straddles the radio jet. In fact, the overall
UV appearance of MO now looks very much like the radio-aligned
rest-frame UV structure seen in the $z = 3.8$ radio galaxy 4C\,41.17
\citep{wvb:99}, which is also thought to be due to jet-induced star
formation \citep{dey:97,bicknell:00}.  This an important qualitative
result because it seems now more appropriate to assume that the star
formation in MO occurs in the shear layers at the wall of the jet,
\ie\ not in a head-on jet-cloud collision. This means that the jet /
cloud collision in MO may be a more complex astrophysical problem than
initially thought.

The jet speed in such a shear layer will be lower than at the center,
due to jet expansion and entrainment \citep{laing:02}. The pressure in
the wall shock is also lower than in the bow-shock of the jet. This
translates to a lower speed for the radiative shocks in the entrained
clouds, and the parameter space studied by \citet{mo_sim} then
indicates that star formation may be easier to accomplish, at least
with a certain range of pre-shock cloud densities. But perhaps more
important is that shear layers of FR-I type jets are active sites of
relativistic particle acceleration as evinced by radio spectral
tomographic imaging which shows a remarkably flat radio spectral energy
distribution across the jets \citep{young:05}. This is further
supported by recent X-ray observations of the nearest FR-I radio jet,
in Centaurus A \citep{kataoka:06}, which show that the jet is
edge-brightened with a possible hardening of the X-ray spectrum in the
outer regions. As a result the pressure of the relativistic fluid in
these regions will be enhanced, which may affect the interaction with
the clouds and the star formation process.

The peak \HI\ column densities in the \HI\ blobs are about $N_{HI} = 4.5
\times 10^{20}$ cm$^{-2}$. This is just at the bottom of the range where, under
galactic conditions, star formation would occur \citep{schaye:04},
but considering the relatively low spatial resolution of the \HI\ data,
it could match more or less. However, the observed star formation
sites are several kpc away from where the peaks in the \HI\ data are,
\ie, the star formation is not happening where the \HI\ says it should
happen.  So it may not be obvious to make the usual link between star
formation and \HI\ column density.  The same is true in the \HI\
filament near Centaurus~A where the \HI\ column densities are well
above $10^{21}$  cm$^{-2}$ but only at those locations where there is {\em no} star formation.

One can also compare the global star formation efficiency 
$M_{stars} / M_{HI}$ in MO to that of other systems. For MO we find 
 $\sim$ 4\%, which is similar to that in Centaurus~A \citep[a few percent;][]{cena}, and indeed typical for starbursts in isolated dwarf galaxies \citep[][and references therein]{fritze:05}.

It is puzzling that, while the \HI\ emission downstream from MO-South
is comparable to that of MO-North, the \ha\ and UV emission and hence
the current star formation activity is much less there.  It should be
noted in this regard that the jet is disrupted in the region of MO, so
it may well be that the jet-cloud interaction is simply stronger in
MO-North than in MO-South. This would be the case if, for example, the
sense of the jet precession was towards the north.

\begin{figure*}
\centering
\includegraphics[width=0.50\linewidth]{f15a.eps}\hspace{0.05\linewidth}
\includegraphics[width=0.40\linewidth]{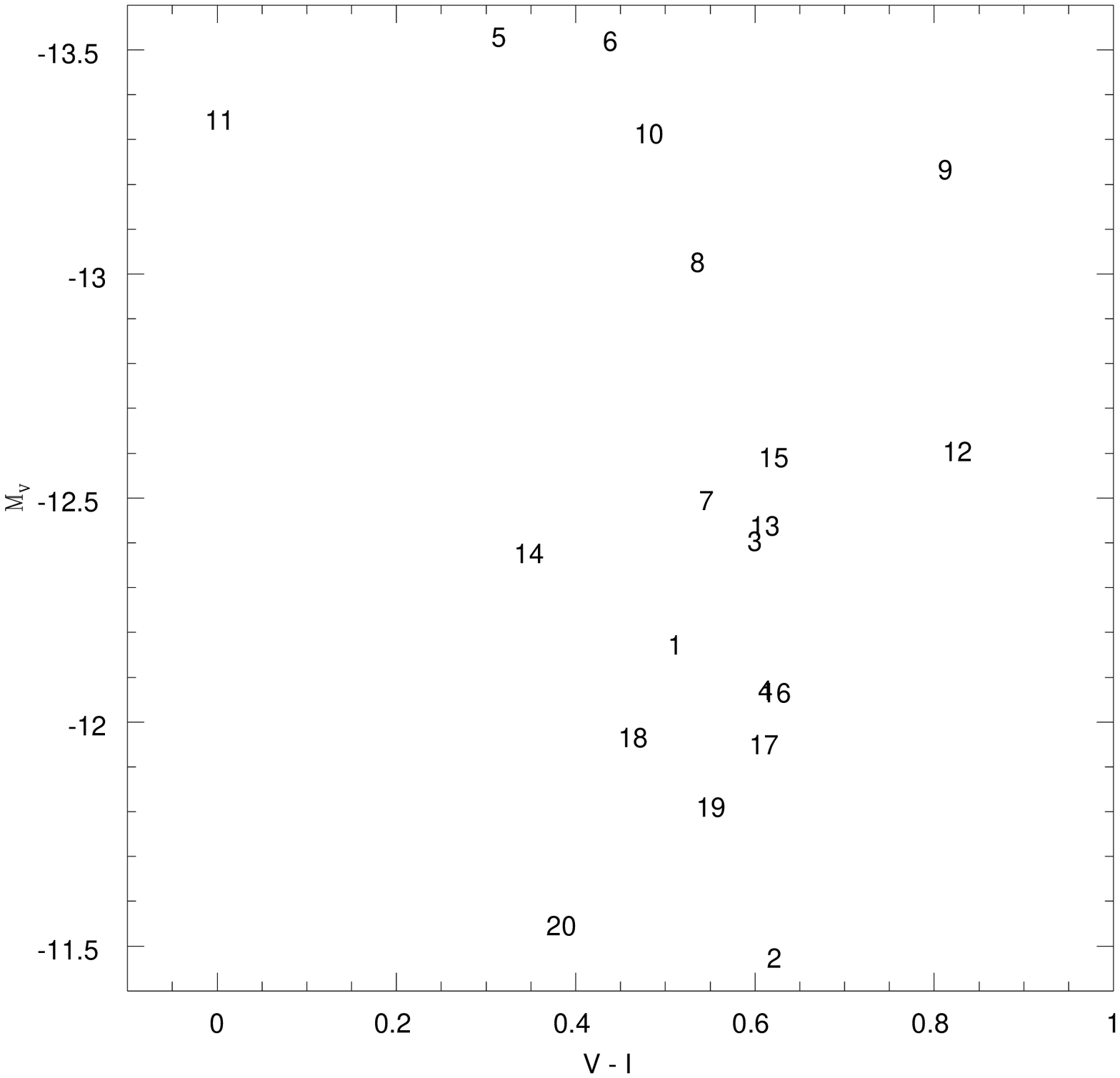}
\caption{\label{fig:blobs} {\em Left:} F555W image of MO, showing the
apertures used for F555W/F814W photometry of ``blobs'', numbered sequentially from south to north. {\em Right:}
Photometry of the ``blobs'' in MO, converted from F555W/F814W to
Johnson $V$ and $I$, and to absolute magnitudes in $V$. Colors and
magnitudes are typical for young star clusters.  }
\end{figure*}

The HST images for the first time reveal the presence of numerous
``blobs'' which appear to be young clusters associated with \HII\
regions. We extracted photometry in 0.5\arcsec-radius apertures for 20
of these blobs (Fig.~\ref{fig:blobs}), converting from the HST filters
to Johnson $V$ and $I$, and transforming the $V$-band measurements to
absolute magnitudes. Comparison with \citet{vdb:95} shows that the
absolute magnitudes are what would be expected of young star clusters,
and the colors of these blobs, when compared to Fig.~11 of
\citet{whitmore:95}, are just what would be expected for ages of a few
Myr (consistent with all our other data). $U$-band data would be
necessary in order to pin down these ages more accurately, and we would need
to reach fainter magnitudes in order to compare with the color-magnitude
diagrams for individual stars in Centaurus~A by \citet{mould:00}.

\clearpage
\clearpage

\subsection{Timing}
\label{sec:timing}

Due to the tendency of a young population to dominate the integrated
starlight in a galaxy, it is difficult to completely rule out the
presence of an older pre-existing population. The situation in MO is
even more complex due to its location in the outskirts of the
elliptical galaxy NGC\,541, and in the bridge of stars connecting
NGC\,541 with two of the other ellipticals in the cluster. Certainly,
the spectroscopic and photometric data presented here agree that a
population of order 7.5\,Myr old is present in MO, and dominates the
light output. The consistency of the pixel ages, the low internal
reddening, and the lack of absorption lines expected from an old
population, also support the conjecture that MO consists of a
single-age population. Additionally, the similar morphologies at F555W
and \kp\ are not what would be expected if an old, gas-rich dwarf
galaxy wandered into the jet and had a new burst of star formation
``switched on'' by the jet-cloud interaction. The star formation also
appears at the leading edge of the \hi\ cloud, as would be expected if
it was the jet that caused the \hi\ and subsequently the stars to
form (since it takes time for the stars to form after the passage of the jet). As discussed above, however, it is not possible using these data
to completely rule out the presence of an older underlying population
in MO. It does seem clear, however, that the picture of a jet-cloud
interaction, sometime a little more than 7.5\,Myr ago, would be
capable of causing the observed phenomena.

Examination of the radio maps shows that a jet plume \citep{wvb:85}
extends up to 37\,kpc ``downstream'' of MO (or somewhat further if the
jet is not in the plane of the sky). It seems unlikely that this is a
buoyant bubble \citep{bk:01} as the jet is still ``switched on'' and
the plume does not display the characteristic ``mushroom cloud''
shape. The fact that the jet is well-collimated until its interaction
with MO, and disrupted downstream, as well as the similar size of the
\hi\ cloud to the width of the jet, suggest that as the jet expanded,
it ran into the gas cloud that ultimately formed MO. If MO was a
pre-existing galaxy that wandered into the jet, it is unlikely that it
would have done so at the precise time to cause the decollimation
seen.

The age of the downstream cocoon can therefore be estimated as the age
of MO, plus the time taken for stars to begin forming after the
jet-cloud interaction. Taking a typical star formation timescale
$\lesssim$~1\,Myr \citep{mo_sim}, we assume that the cocoon expanded
by at least 37\,kpc in 8.5\,Myr, \ie, at $\gtrsim 4 \times
10^3$\,\kms.

If we assume that the SFR has been roughly constant during the
lifetime of MO, this yields a timescale for the formation of the $1.9
\times 10^7$\,\msun\ of stars of 36.5\,Myr. This is only a few times the 7.5\,Myr inferred from the SED age dating. Since it is plausible that the SFR was somewhat higher in the past, it is reasonable to assume that all the stars may have been formed since the jet-cloud interaction. The timescale for depletion of the observed \hi\ is much longer ($\sim 1$\,Gyr).

\section{Conclusions}

We have presented new evidence that the star formation in MO was
induced by a radio jet, primarily in the outer shear layers. This is
similar to the jet-induced star formation associated with the Centaurus
A jet, and the radio-aligned star forming regions in powerful radio
galaxies at high redshift. Absorption and emission line measurements,
and broad-band SED fitting, give a consistent age of around 7.5\,Myr
for MO. While it is not possible to completely rule out the presence
of an old population in MO, the data are consistent with MO having
formed {\em de novo} when the jet interacted with the ambient ISM /
IGM.  We discovered a double neutral hydrogen cloud which straddles
the radio jet downstream of MO, at a location where the jet changes
direction and de-collimates. Unlike Centaurus A, we propose that the \HI\ associated with MO
formed {\it in situ} through cooling of clumpy warm gas, in the
stellar bridge or cluster IGM, as it was compressed by radiative shocks
at the jet collision site, in agreement with earlier numerical
simulations \citep{mo_sim}. The star formation in MO then followed
from the cooling and collapse of such \HI\ clouds, and the \HI\
kinematics, which show 40\,\kms\ shear velocites, are also consistent
with such models. The FR-I type radio source associated with MO is
orders of magnitude lower in radio luminosity, and correspondingly
more common, than the very luminous FR-II type radio galaxies and
their aligned star forming regions at high redshift. Nevertheless, as
MO shows, also these types of radio sources can trigger star
formation, even if the ambient medium is only moderately dense. This
suggests that jet-induced star formation may be quite common in the
early universe, where gas densities in their parent, forming galaxies
and surrounding proto-clusters were much higher than today, and when
AGN activity was much more prevalent.

\acknowledgments

We thank the staff of the Lick and Siding Spring Observatories for
their support. Thanks to Michael Gregg for providing us with Lick
PFCam data. We thank the anonymous referee for helpful and constructive
suggestions. Data presented herein were obtained at the W.\ M.\ Keck
Observatory, which is operated as a scientific partnership among the
California Institute of Technology, the University of California and
the National Aeronautics and Space Administration. The Observatory was
made possible by the generous financial support of the W.\ M.\ Keck
Foundation. Work was also based on observations made with the NASA/ESA
Hubble Space Telescope, obtained from the data archive at the Space
Telescope Science Institute. STScI is operated by the Association of
Universities for Research in Astronomy, Inc.\ under NASA contract NAS
5-26555. GALEX (Galaxy Evolution Explorer) is a NASA Small Explorer,
launched in April 2003. We gratefully acknowledge NASA's support for
construction, operation, and science analysis for the GALEX mission,
developed in cooperation with the Centre National d'Etudes Spatiales
of France and the Korean Ministry of Science and Technology. The
National Radio Astronomy Observatory is a facility of the National
Science Foundation operated under cooperative agreement by Associated
Universities, Inc. Some of the data presented in this paper were
obtained from the Multimission Archive at the Space Telescope Science
Institute (MAST). STScI is operated by the Association of Universities
for Research in Astronomy, Inc., under NASA contract
NAS5-26555. Support for MAST for non-HST data is provided by the NASA
Office of Space Science via grant NAG5-7584 and by other grants and
contracts. Dopita acknowledges the support of both the Australian
National University and the Australian Research Council (ARC) through
his ARC Australian Federation Fellowship and of financial support
through ARC Discovery project grant DP0208445. Work was performed
under the auspices of the U.~S.\ Department of Energy, National
Nuclear Security Administration by the University of California,
Lawrence Livermore National Laboratory under contract No.\
W-7405-Eng-48. SC and WvB acknowledge support for radio galaxy studies at 
UC Merced, including the work reported here, with the Hubble, Spitzer and 
Chandra space telescopes via NASA grants HST \#10127, SST \#3482, SST \#3329 
and Chandra / CXO \#06701011.


\begin{thebibliography}{}

\bibitem[Anninos, Fragile, \& Murray(2003)]{cosmos} Anninos, P., Fragile, P.~C., \& Murray, S.~D.\ 2003, \apjs, 147, 177 

\bibitem[Avni(1976)]{avni:76} Avni, Y.\ 1976, \apj, 210, 642 

\bibitem[Bicknell et al.(2000)]{bicknell:00} Bicknell, G.~V., Sutherland, R.~S., van Breugel, W.~J.~M., Dopita, M.~A., Dey, A., \& Miley, G.~K.\ 2000, \apj, 540, 678 

\bibitem[B{\" o}ker et al.(1999)]{boker:99} B{\" o}ker, T., et al.\ 1999, \apjs, 124, 95

\bibitem[Brodie et al.(1983)]{brodie:83} Brodie, J., Koenigl, A., \& Bowyer, S.\ 1983, \apj, 273, 154 

\bibitem[Br{\"u}ggen \& Kaiser(2001)]{bk:01} Br{\"u}ggen, M., \& Kaiser, C.~R.\ 2001, \mnras, 325, 676 

\bibitem[Calzetti et al.(1994)]{cal:94} Calzetti, D., Kinney, A.~L., \& Storchi-Bergmann, T.\ 1994, \apj, 429, 582 

\bibitem[Calzetti(2001)]{cal:01} Calzetti, D.\ 2001, New Astronomy Review, 45, 601 

\bibitem[Chambers, Miley, \& van Breugel(1987)]{cmvb:87} Chambers, K.~C., Miley, G.~K., \& van Breugel, W.\ 1987, \nat, 329, 604 

\bibitem[de Robertis et al.(1987)]{derob:87} de Robertis, M.~M., Dufour, R.~J., \& Hunt, R.~W.\ 1987, \jrasc, 81, 195

\bibitem[De Young(1989)]{deyoung:89} De Young, D.~S.\ 1989, \apjl, 342, L59

\bibitem[Dey et al.(1997)]{dey:97} Dey, A., van Breugel, W., Vacca, W.~D., \& Antonucci, R.\ 1997, \apj, 490, 698 

\bibitem[Dopita et al.(1984)]{dopita:84} Dopita, M.~A., Binette, 
L., Dodorico, S., \& Benvenuti, P.\ 1984, \apj, 276, 653 

\bibitem[Dopita \& Ryder(1994)]{dopita:94}  Dopita, M.~A. \& Ryder,S.~D. 1994, \apj, 430, 163

\bibitem[Dopita \& Sutherland(1995)]{ds:95} Dopita, M.~A., \& 
Sutherland, R.~S.\ 1995, \apj, 455, 468

\bibitem[Dopita \& Hua(1997)]{dopita:97} Dopita, M.~A., \& Hua, C.~T.\ 1997, \apjs, 108, 515

\bibitem[Dopita et al.(2000)]{dopita:00} Dopita, M.~A., Kewley, L.~J., Heisler, C.~A., \& Sutherland, R.~S.\ 2000, \apj, 542, 224 

\bibitem[Dopita et al.(2002)]{dopita:02} Dopita, M.~A., Groves, B.~A., Sutherland, R.~S., Binette, L., \& Cecil, G.\ 2002, \apj, 572, 753

\bibitem[Eilek et al.(2002)]{eilek:02} Eilek, J., Hardee, P., Markovic, T., Ledlow, M., \& Owen, F.\ 2002, New Astronomy Review, 46, 327

\bibitem[Fanaroff \& Riley(1974)]{fr} Fanaroff, B.~L., \& Riley, J.~M.\ 1974, \mnras, 167, 31P 

\bibitem[Ferland et al.(2002)]{ferland:02} Ferland, G.~J., Fabian, A.~C., \& Johnstone, R.~M.\ 2002, \mnras, 333, 876 

\bibitem[Filippenko(1982)]{difrefrac} Filippenko, A.~V.\ 1982, \pasp, 94, 715

\bibitem[Fragile et al.(2004)]{mo_sim} Fragile, P.~C., Murray, S.~D., Anninos, P., \& van Breugel, W.\ 2004, \apj, 604, 74 

\bibitem[Fritze-v.~Alvensleben(2005)]{fritze:05} Fritze-v.~Alvensleben, U.\ 2005, preprint (astro-ph/0508099)

\bibitem[Gil de Paz et al.(2006)]{galexngs} Gil de Paz, A., et al.\ 2006, in prep.

\bibitem[Graham \& Price(1981)]{graham:81} Graham, J.~A., \& Price, R.~M.\ 1981, \apj, 247, 813 

\bibitem[Gonz{\'a}lez Delgado et al.(1999)]{gd:99} Gonz{\'a}lez Delgado, R.~M., Leitherer, C., \& Heckman, T.~M.\ 1999, \apjs, 125, 489 

\bibitem[Johnston et al.(2005)]{johnston:05} Johnston, H.~M., Hunstead, R.~W., Cotter, G., \& Sadler, E.~M.\ 2005, \mnras, 356, 515 

\bibitem[Kataoka et al.(2006)]{kataoka:06} Kataoka, J., Stawarz, L., Aharonian, F., Takahara, F., Ostrowski, M., \& and Edwards, P.~G.\ 2006, \apj, in press

\bibitem[Kennicutt(1998)]{kennicutt:98} Kennicutt, R.~C.\ 1998, \araa, 36, 189

\bibitem[Kewley \& Dopita(2002)]{kewley:02} Kewley, L.~J., \& Dopita, M.~A.\ 2002, \apjs, 142, 35

\bibitem[Klamer et al.(2004)]{klamer:04} Klamer, I.~J., Ekers, R.~D., Sadler, E.~M., \& Hunstead, R.~W.\ 2004, \apjl, 612, L97 

\bibitem[Klein, McKee, \& Colella(1994)]{klein:94} Klein, R.~I., McKee, C.~F., \& Colella, P.\ 1994, \apj, 420, 213 

\bibitem[Laing \& Bridle(2002)]{laing:02} Laing, R.~A., \& Bridle, A.~H.\ 2002, \mnras, 336, 1161 

\bibitem[Leitherer et al.(1999)]{sb99} Leitherer, C., et al.\ 1999, \apjs, 123, 3

\bibitem[Martin et al.(2005)]{galex} Martin, D.~C., et al.\ 2005, \apjl, 619, L1

\bibitem[Minkowski(1958)]{mink} Minkowski, R.\ 1958, \pasp, 70, 143 

\bibitem[McNamara(2002)]{mcnamara:02} McNamara, B.~R.\ 2002, New Astronomy Review, 46, 141
 
\bibitem[Morganti et al.(2005)]{morganti:05} Morganti, R., Tadhunter, C.~N., \& Oosterloo, T.~A.\ 2005, \aap, 444, L9 

\bibitem[Morrissey et al.(2005)]{galexpipe} Morrissey, P., et al.\ 2005, \apjl, 619, L7

\bibitem[Mould et al.(2000)]{mould:00} Mould, J.~R., et al.\ 2000, \apj, 536, 266

\bibitem[Nikogossyan et al.(1999)]{a194} Nikogossyan, E., Durret, F., Gerbal, D., \& Magnard, F.\ 1999, \aap, 349, 97 

\bibitem[O'Dea et al.(2004)]{odea:04} O'Dea, C.~P., Baum, S.~A., Mack, J., Koekemoer, A.~M., \& Laor, A.\ 2004, \apj, 612, 131 

\bibitem[Oke et al.(1995)]{lris} Oke, J.~B., et al.\ 1995, \pasp, 107, 375

\bibitem[Oosterloo \& Morganti(2005)]{cena} Oosterloo, T.~A., \& Morganti, R.\ 2005, \aap, 429, 469 

\bibitem[Panuzzo et al.(2003)]{panuzzo:03} Panuzzo, P., Bressan, A., Granato, G.~L., 
Silva, L. \& Danese, L. 2003, \aap, 409, 99

\bibitem[Rees(1989)]{rees:89} Rees, M.~J.\ 1989, \mnras, 239, 1P 

\bibitem[Schaye(2004)]{schaye:04} Schaye, J.\ 2004, \apj, 609, 667 

\bibitem[Schiminovich et al.(1994)]{schim:94} Schiminovich, D., van Gorkom, J.~H., van der Hulst, J.~M., \& Kasow, S.\ 1994, \apjl, 423, L101 

\bibitem[Schlegel et al.(1998)]{galred} Schlegel, D.~J., Finkbeiner, D.~P., \& Davis, M.\ 1998, \apj, 500, 525

\bibitem[Sheinis et al.(2002)]{esi} Sheinis, A.~I., Bolte, M., Epps, H.~W., Kibrick, R.~I., Miller, J.~S., Radovan, M.~V., Bigelow, B.~C., \& Sutin, B.~M.\ 2002, \pasp, 114, 851

\bibitem[Simkin(1976)]{simkin:76} Simkin, S.~M.\ 1976, \apj, 204, 251 

\bibitem[Smith et al.(2000)]{a194z} Smith, R.~J., Lucey, J.~R., Hudson, M.~J., Schlegel, D.~J., \& Davies, R.~L.\ 2000, \mnras, 313, 469 

\bibitem[Spergel et al.(2003)]{wmap} Spergel, D.~N., et al.\ 2003, \apjs, 148, 175

\bibitem[Stasinska \& Leitherer(1996)]{sl:96} Stasinska, G., \& Leitherer, C.\ 1996, \apjs, 107, 661 

\bibitem[Sutherland \& Dopita(1993)]{sd:93} Sutherland, R.~S., \& Dopita, M.~A.\ 1993, \apjs, 88, 253 

\bibitem[Trager et al.(2000)]{n547z} Trager, S.~C., Faber, S.~M., Worthey, G., \& Gonz{\'a}lez, J.~J.\ 2000, \aj, 119, 1645 

\bibitem[van Breugel et al.(1985)]{wvb:85} van Breugel, W., Filippenko, A.~V., Heckman, T., \& Miley, G.\ 1985, \apj, 293, 83 

\bibitem[van Breugel et al.(1999)]{wvb:99} van Breugel, W., Stanford, A., Dey, A., Miley, G., Stern, D., Spinrad, H., Graham, J., \& McCarthy, P.\ 1999, The Most Distant Radio Galaxies, 49 

\bibitem[van den Bergh(1995)]{vdb:95} van den Bergh, S.\ 1995, \nat, 374, 215 

\bibitem[Veilleux \& Osterbrock(1987)]{vo:87} Veilleux, S., \& Osterbrock, D.~E.\ 1987, \apjs, 63, 295 

\bibitem[Verdoes Kleijn et al.(1999)]{mo:hst} Verdoes Kleijn, G.~A., Baum, S.~A., de Zeeuw, P.~T., \& O'Dea, C.~P.\ 1999, \aj, 118, 2592

\bibitem[Whitmore \& Schweizer(1995)]{whitmore:95} Whitmore, B.~C., \& Schweizer, F.\ 1995, \aj, 109, 960 

\bibitem[Wills et al.(2004)]{wills:04} Wills, K.~A., Morganti, R., Tadhunter, C.~N., Robinson, T.~G., \& Villar-Martin, M.\ 2004, \mnras, 347, 771 

\bibitem[Young et al.(2005)]{young:05} Young, A., Rudnick, L., Katz, D., DeLaney, T., Kassim, N.~E., \& Makishima, K.\ 2005, \apj, 626, 748

\end{thebibliography}
\end{document}